\documentstyle[prd,aps,eqsecnum,preprint,tighten,floats,epsf]{revtex}
\def\senk#1{\bbox{#1}_\perp}
\def\ha{{1\over 2}}
\def\ub#1{\underline{#1}}
\def\ths{\thinspace}
\def\psibar{\overline{\psi}}
\def\del{\partial}
\def\ra{\rightarrow}
\def\eg{{\it e.g.}}
\def\g{\gamma}

\begin{document}
\draft

\preprint{\vbox{\hfill SLAC-PUB-7745 \\
          \vbox{\hfill UMN-D-98-1}  \\
          \vbox{\hfill SMUHEP/98-01 }
          \vbox{\vskip0.3in}
          }}

\title{Pauli--Villars as a Nonperturbative
Ultraviolet Regulator in 
Discretized Light-Cone Quantization}

\author{Stanley J. Brodsky%
\footnote{\baselineskip=14pt
Work supported in part by the Department of Energy,
contract DE-AC03-76SF00515.}}
\address{Stanford Linear Accelerator Center,
Stanford University, Stanford, California 94309}

\author{John R. Hiller}
\address{Department of Physics,
University of Minnesota, Duluth, Minnesota 55812}

\author{Gary McCartor%
\footnote{\baselineskip=14pt
Work supported in part by the Department of Energy,
contract DE-FG03-95ER40908.}}
\address{Department of Physics,
Southern Methodist University, Dallas, Texas 75275}

\date{\today}

\maketitle

\begin{abstract}
We propose a solution to the problem of renormalizing
light-cone Hamiltonian theories while maintaining
Lorentz invariance and other symmetries.  The method
uses generalized Pauli--Villars regulators to render the theory
finite.  We discuss the method in the context of Yukawa
theory at one loop and for a soluble model in $3+1$
dimensions.  The model is studied nonperturbatively.
Numerical results obtained with use of discretized
light-cone quantization, special integration weighting
factors, and the complex symmetric Lanczos diagonalization
algorithm compare well with the analytic answers.
\end{abstract}
\pacs{12.38.Lg,11.15.Tk,11.10.Gh,02.60.Nm
\begin{center}(Submitted to Physical Review D.)\end{center}}

\narrowtext

\section{Introduction}

DLCQ (Discretized Light-Cone
Quantization)\cite{PauliBrodsky,DLCQreview,RecentDLCQ}
is a suggested computational procedure in which one
specifies quantization conditions on the characteristic 
surface $x^+\equiv (x^0 + x^1) = 0$, introduces
periodicity conditions to induce a
discrete basis, truncates the basis set by some procedure 
to produce a finite matrix, then takes the spectrum and
eigenvectors of that matrix as an approximation to the
physical spectrum and wave functions.  The difficulty is 
that, as always in quantum field 
theory, removing an infinite set of high (bare) energy 
states induces a renormalization of the operators.  A 
completely consistent procedure for performing the 
truncation and renormalization has not yet been demonstrated, but the
problem has received considerable study.  Some calculations 
have been published\cite{SomeCalculations}
which simply use the periodicity conditions
combined with a momentum cutoff.  While the numerical results
are accurate for superrenormalizable theories such as
$1+1$-dimensional gauge theories, it is clear that
such a procedure is problematical for renormalizable theories
in $3+1$ dimensions.   A systematic renormalization procedure
has been proposed by Wilson, Perry and co-workers\cite{Wilson}.
These authors use a cutoff chosen for
computational convenience and then try to find the mixing of the operators
under renormalization using ideas of the Wilson renormalization group.
Since the procedure makes no attempt to preserve the symmetries of the
theory, one expects a great deal of mixing and many counterterms; some
skill in guessing the counterterms seems to be required.

In this paper we will suggest a procedure which lies somewhat closer to
traditional ideas in field theory than the Wilson plan in that we will
make an attempt to preserve more of the symmetries of the theory; but
the objective is the same.  The idea is to add enough Pauli--Villars
fields\cite{PauliVillars} to the theory to regulate perturbation theory.
Having done that
we hope that, since the theory is basically finite, the periodicity
conditions and momentum cutoff will be sufficiently benign to allow
a consistent renormalization to be performed.  In the Wilson language,
we hope that the heavy fields will add the necessary counterterms
automatically with no particular cleverness from us.

In the next section we consider the one-loop fermion self mass in
Yukawa theory \cite{YukawaLFTD}.
This problem has been considered previously in the
literature\cite{PreviousYukawa}, but we shall discuss the analysis in
the context of using the Pauli--Villars program to preserve the
discrete chiral symmetry of the theory. The computation requires three
Pauli--Villars fields for proper regularization, including the
elimination of all terms --- including spurious finite terms ---  not
proportional to the bare fermion mass squared.

We then present and solve a model field theory very similar to the
scalar field model studied in the 1950's by Greenberg and
Schweber\cite{SchweberGreenberg}.  This model, which requires
renormalization,
allows us to illustrate the procedure and to examine some important
numerical issues, at least within the context of the model.  These
issues include the number of states which must be devoted to the heavy
fields and the related question of how heavy their masses must be.
Section~\ref{sec:DLCQapplied} discusses the numerical solution of this
same model and compares these results with the analytic solution.

A final section contains our general conclusions.  This is followed by
two appendices that provide details of our light-cone conventions
and of improved methods for accurate DLCQ calculations.

\section{Regularization of the Fermion Self-Energy
in Light-Cone Quantization}

\subsection{Analysis}

We consider Yukawa theory defined by the action
\begin{equation}
S=\int d^4x \Biggl[\ha(\del_\mu\phi)^2-\ha\mu^2\phi^2
	+{i\over2}\Bigl(\psibar\g^\mu\del_\mu-(\del_\mu\psibar)\g^\mu\Bigr)\psi
		-M\psibar\psi - g\phi\psibar\psi-\lambda\phi^4\Biggr].
\end{equation} 
For the problem of interest here, the $\lambda\phi^4$ interaction will
not be needed.  The operator $P^-$ which controls the dynamics is
\begin{equation}
P^-=\ha\int dx^-d^2x \ths\ths T^{+-}\,,
\end{equation}
where
\begin{eqnarray}
T^{+-}&=&(\bbox{\del}_\perp\phi)^2+ \mu^2\phi^2 -i\psi_-^\dagger(\del^+\psi_-)
\nonumber\\
&\qquad&+2\psi_-^\dagger \g^0 (-i\g^i \del_i +M+g\phi)\psi_+
+ {\rm h.c.}\ths
\end{eqnarray}
The field $\psi_-$ is nondynamical and can be eliminated via the constraint
relation
\begin{equation}
i\del_-\psi_-=\ha\g^0\Bigl[ -i\g^i\del_{i} + M + g\phi\Bigr]\psi_+ .
\end{equation} 
For the second order shift in the eigenvalue of the operator $P^-$ 
of the one-fermion state  with $\bbox{p}_\perp=0$, one easily calculates
\begin{equation}
  \frac{\alpha}{2\pi^2} \int_0^1\frac{dx}{1-x}
     \int d^2q_\perp
     \frac{\bbox{q}_\perp^2+(2-x)^2M^2}
           {\bbox{q}_\perp^2+x^2M^2+(1-x)\mu^2}\,,
\end{equation}
where
\begin{equation}
     \alpha \equiv \frac{g^2}{4\pi}\,.
\end{equation}

The integral is divergent in the ultraviolet and must be regulated.
Let us first
consider regulating the integral with a momentum cutoff.  While 
several possibilities might be considered, including a cutoff on 
$\senk{q}$ alone, the most commonly used cutoffs couple $q^+$
and $\senk{q}$ in some way.  To retain boost invariance we
will consider the ``invariant
mass'' cutoff in which the total invariant mass
of the intermediate state is limited\cite{LepageBrodsky}.
For the present case this rule gives
\begin{equation}
      \frac{q_\perp^2 + \mu^2}{x} +
          \frac{q_\perp^2 + M^2}{1-x} \le \Lambda^2\,.
\end{equation}
This cutoff also appears if we simply limit the
change in mass of the matrix elements of
the interaction Hamiltonian.
For the integral we then get
\widetext
\begin{equation}
I(\mu^2,M^2,\Lambda^2)\equiv
     \frac{\alpha}{2\pi^2}\int_{L_-}^{L_+}\frac{dx}{1-x}
     \int_{\bbox{q}_\perp^2\leq L}d^2q_\perp
     \frac{\bbox{q}_\perp^2+(2-x)^2M^2}
           {\bbox{q}_\perp^2+x^2M^2+(1-x)\mu^2}\,,
\end{equation}
where
\begin{eqnarray}
L_{\pm}&=&\frac{1}{2\Lambda^2}\left[\Lambda^2+\mu^2-M^2
\pm\sqrt{(\Lambda^2+\mu^2-M^2)^2-4\Lambda^2\mu^2}\right]\,,
\nonumber \\
L&=&\Lambda^2x(1-x)-\mu^2(1-x)-M^2x\,.
\end{eqnarray}
\narrowtext
\noindent
The integral can be done in closed form, but the result for
arbitrary parameters is not terribly
illuminating.  If we take $\Lambda^2\gg \mu^2\gg M^2$ we get
\begin{eqnarray}  \label{eq:Iapprox}
I(\mu^2,M^2,\Lambda^2)&\approx& \frac{\alpha}{2\pi}
     \Biggl[ \left( \frac{\Lambda^2}{2} - \mu^2 \ln \Lambda^2 
           + \mu^2 \ln \mu^2 - \frac{\mu^4}{2\Lambda^2}\right)
  \nonumber \\
&& + M^2\left( 3 \ln\Lambda^2 - 3 \ln \mu^2 - \frac{9}{2}
                           + \frac{5\mu^2}{\Lambda^2}\right)
   \\
&& + M^4\left(\frac{2}{\mu^2}\ln (M^2/\mu^2)
                 +\frac{1}{3\mu^2}-\frac{1}{2\Lambda^2}\right)
                           \Biggr]\,.
\nonumber
\end{eqnarray}
Perhaps the most striking thing about this result is not
so much that it is quadratically divergent as
$\Lambda\rightarrow\infty$, but that it
does not go to zero with $M$.  This is in contrast to
the Feynman result.  In fact the vanishing of the self 
mass with vanishing bare mass
is formally protected by the discrete chiral symmetry: 
$\psi\ra i\gamma_5\psi,$ with $\phi\ra-\phi$.
That $I$ is not proportional to $M^2$ is due to the fact that
the invariant mass regulator does not preserve this
symmetry.

It is well known that the four-dimensional Feynman integral
for the fermion self energy can be regulated by the
addition of one Pauli--Villars boson field.  If that is done
the integral is then ``finite" by power counting and vanishes with $M$.
One might then think that if one first performs the $q^-$
integral one would get a finite three-dimensional light-cone
integral.  However, the Feynman integral is only conditionally convergent,
and therefore any value ascribed to it is a
prescription.  The standard integration procedure - symmetric
integration in the spatial momenta with the $q^0$ integral done last -
preserves the discrete chiral symmetry and
thus leads to the vanishing of the result at $M = 0$.
The terms in (\ref{eq:Iapprox}) which
do not vanish as $M\rightarrow0$ at large $\Lambda$
include terms quadratic in $\Lambda$, logarithmic
in $\Lambda$, and independent of $\Lambda$.
Therefore three Pauli--Villars bosonic fields are necessary
to render the light-cone integral consistent with
discrete chiral symmetry. The entire light-cone
integral (\ref{eq:Iapprox}) is then finite and vanishes
as $M\rightarrow0$.  Thus we need three Pauli--Villars
conditions:
\begin{equation}
\alpha+\sum_{i=1}^3 \alpha_i=0\,, \;\;
\alpha\mu^2+\sum_{i=1}^3 \alpha_i\mu_i^2=0\,, \;\;
\sum_{i=1}^3 \alpha_i\mu_i^2\ln(\mu_i^2/\mu^2)=0\,,
\end{equation}
where the $\alpha_i$'s and $\mu_i$'s are the coupling
constants and masses of the heavy fields.
The logarithmic divergent term $3M^2\ln\Lambda^2$
in (\ref{eq:Iapprox}) returns if the masses of the heavy fields
go to infinity, but the nonzero value at $M = 0$ does
not return.  The fact that three Pauli--Villars fields are necessary to
regulate the self-energy graph in the light-cone
representation is an old result\cite{ChangYan}, and
it has received considerable study in\cite{BurkardtLangnau}.
One might wonder whether
there is some feature of the theory from a purely 
equal-time perspective that would allow one to predict 
the number of heavy fields necessary to regulate the 
calculation in the light-cone
representation.\footnote{We are
not certain, but we speculate that it may be the same 
number necessary to regulate time-ordered perturbation
theory in the equal-time representation; perhaps the 
results of\cite{RobertsonMcCartor} on
the connection between the two representations could 
then be extended to higher order.}

To perform a DLCQ calculation one must limit
the range of the momenta (which the $\Lambda$ cutoff 
does) as well as make the momenta discrete by
introducing periodicity conditions for the fields on 
the surface $x^+ = 0$.  We may take $\psi_+=\Lambda_+\psi$ to be
antiperiodic in $x^-$ and periodic in $\bbox{x}_\perp$.
(See Appendix~\ref{sec:LCcoordinates}.) The
scalar fields are taken to be periodic in both $x^-$ 
and $\bbox{x}_\perp$. With the heavy fields and the momentum
cutoff in place, the only effect of the periodicity
conditions on the above perturbation calculation is that 
the finite integral is then evaluated as a discrete sum.
The convergence of the discrete sum to the continuum result
is discussed below.

The discussion in this section suggests a new general
procedure for resolution of the UV divergences of
light-cone Hamiltonian field theory.
The heavy fields will produce the counterterms necessary
to make a consistent renormalization possible.
What we propose is to test this
procedure nonperturbatively, that is, include enough heavy 
fields in the Lagrangian to regulate perturbation theory, 
then produce a finite matrix with a momentum cutoff and 
discretization.
The periodicity conditions can also lead to
constrained zero modes, as discussed in\cite{McCartorRobertson}.
Another important
advantage of the Pauli--Villars fields is that the terms
from the constrained zero modes which would affect the
one loop  mass shift (the most singular terms due to the constrained 
zero modes) are zero at the level of $P^-$.  In the present paper we 
shall not attempt a full Yukawa calculation.  We shall illustrate the 
method for a soluble model and provide some 
discussion of numerical issues.

\subsection{Discrete Evaluation of the Light-Cone Integral}

\subsubsection{DLCQ} \label{sec:DLCQ}

From the periodicity conditions in the light-cone box
\begin{equation}
-L<x^-<L\,,\;\; -L_\perp<x,y<L_\perp\,,
\end{equation}
one obtains discrete momenta
\begin{equation}
p^+\rightarrow\frac{\pi}{L}n\,, \;\;
\bbox{p}_\perp\rightarrow
     (\frac{\pi}{L_\perp}n_x,\frac{\pi}{L_\perp}n_y)\,,
\end{equation}
with $n$ even for bosons and odd for fermions.
Integrals are then replaced by discrete sums obtained as 
trapezoidal approximations on the grid of momentum values
\begin{equation} \label{eq:rawDLCQ}
\int dp^+ \int d^2p_\perp f(p^+,\bbox{p}_\perp)\simeq
   \frac{2\pi}{L}\left(\frac{\pi}{L_\perp}\right)^2
   \sum_n\sum_{n_x,n_y=-N_\perp}^{N_\perp}
   f(n\pi/L,\bbox{n}_\perp\pi/L_\perp)\,.
\end{equation}
The limit $L\rightarrow\infty$ can be exchanged for a limit
in terms of the integer {\em resolution}\cite{PauliBrodsky}
$K\equiv\frac{L}{\pi}P^+$.
The longitudinal momentum fraction $x=p^+/P^+$ becomes $n/K$.  
$H_{\rm LC}$ is independent of $L$.

Because the longitudinal integers $n$ are always positive, DLCQ
automatically limits the number of particles to no more than $\sim\!\!K/2$.
The integers $n_x$ and $n_y$ range between limits associated with
some maximum integer $N_\perp$ fixed by $L_\perp$ and a cutoff that 
limits transverse momentum.

The momentum-space continuum limit is reached
when $K$ and $N_\perp$ become infinite.
The transverse length scale $L_\perp$
is chosen such that $N_\perp\pi/L_\perp$ is the
largest transverse momentum allowed by the cutoff.
The integrations for Pauli--Villars subtractions use the
transverse scale $L_\perp$ determined for the physical boson.
This insures use of a common grid that can easily represent
momentum conservation in interactions.

We then compute the dimensionless integral
\begin{equation}
\tilde{I}(\mu^2,M^2,\Lambda^2)\equiv
   \frac{2\pi^2}{\alpha\mu^2}I(\mu^2,M^2,\Lambda^2)
\end{equation}
and the subtracted integral
\begin{equation}
\tilde{I}_{\rm sub}(M^2/\mu^2,\mu_i^2/\mu^2,\Lambda^2/\mu^2)\equiv
  \tilde{I}(\mu^2,M^2,\Lambda^2)+
    \sum_i \frac{\alpha_i}{\alpha}\tilde{I}(\mu_i^2,M^2,\Lambda^2)\,.
\end{equation}
When the loop integrals are evaluated numerically, the proportion
of error increases with each subtraction.  Once all
three Pauli--Villars subtractions are done, the error can dominate.
This is the case for an ordinary DLCQ calculation.  
The individual integrals are large and therefore
must be computed accurately for the differences to
be accurate.  It can be helpful to have well
separated Pauli--Villars masses, because the coefficients $\alpha_i$ are
then ${\cal O}(1)$ and do not amplify the errors.

The domain of integration, as defined by the invariant-mass cutoff, is
not commensurate with the DLCQ grid.  This causes errors of
two types: one is a truncation error where the edge of the
domain is not properly counted and the other is the loss of 
rotational symmetry in the transverse grid.  In one-dimensional
cases, only the former type occurs.  In that context, it
is easily handled as part of an extrapolation in $K$; 
in the context of the three iterated integrals used for
three dimensions, the error becomes much less controlled.

\begin{figure}
\centerline{\epsfxsize=\columnwidth \epsfbox{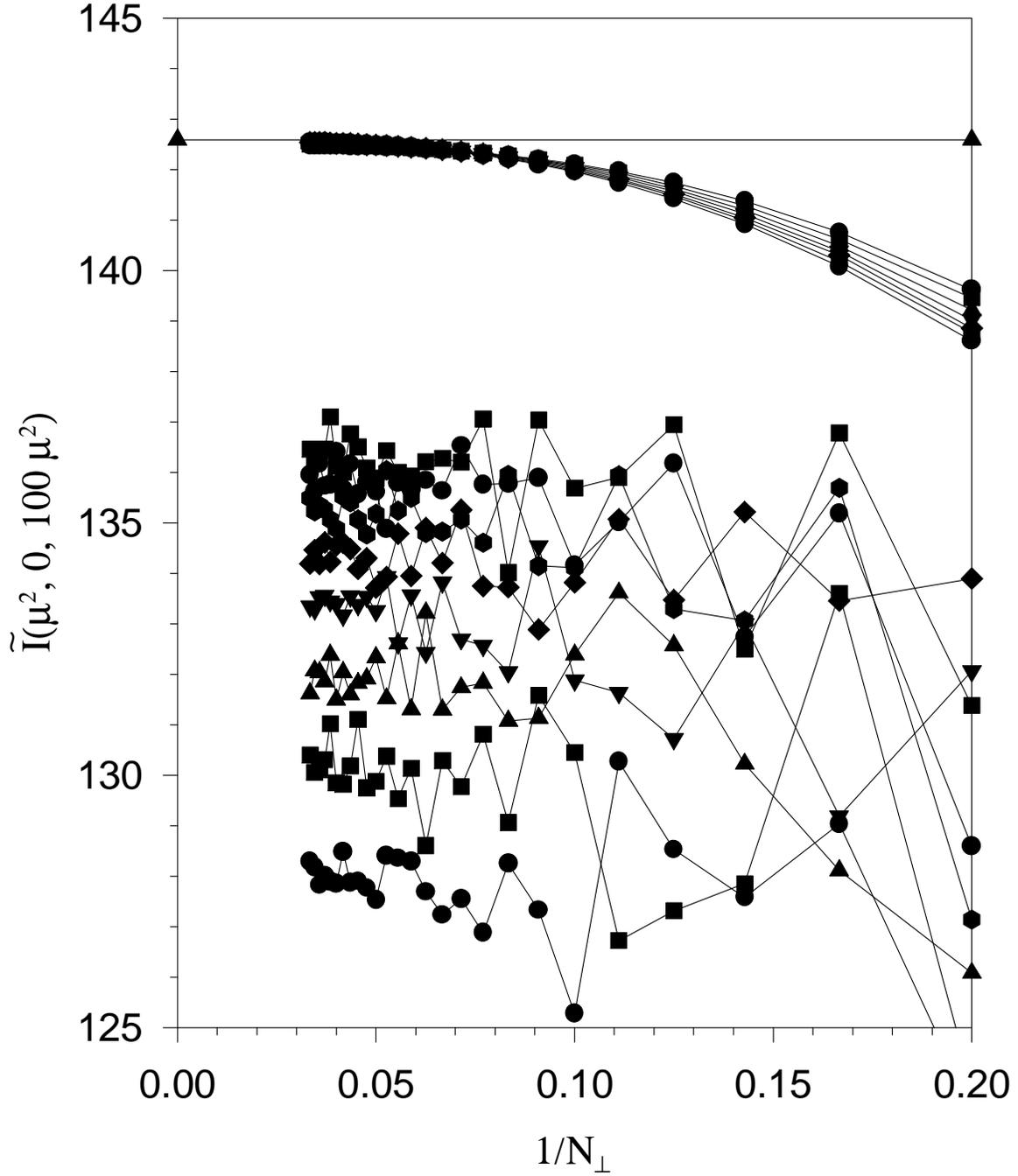} }
\caption{\label{fig:OneLoopSEcirc}
One-loop fermion self energy.
The horizontal line is the exact result.  The smoothly curved results
come from use of transverse circular weighting and longitudinal
Simpson weighting.  The scattered results are from 
ordinary DLCQ calculations.
The DLCQ grid parameters take the ranges $K=10,\,12,\ldots,24$ and
$N_\perp=5,\,6,\ldots,30$.
The lines connect points calculated with the same $K$ value.}
\end{figure}

If the trapezoidal rule is applied to each iterated
integral, as is done in the standard DLCQ approach,
the errors will not follow a systematic dependence
on the grid spacings for a reasonable number of grid points.
This lack of systematic dependence on $K$ and $N_\perp$
can be seen in Fig.~\ref{fig:OneLoopSEcirc}.

These errors could be overcome with a commensurate grid
that uses polar coordinates in the transverse direction.
This would not be easily extended to situations with 
more than two particles.  Also, it turns out that although
a commensurate grid controls the errors in a systematic way,
the errors are still large.  Other methods that use the
DLCQ grid have been found superior.

Given the rectangular DLCQ grid, one can improve on the
simple application of the trapezoidal rule used in (\ref{eq:rawDLCQ}).
The alternative integration schemes \cite{Luchini} are of the general form
\begin{equation}
\int d^nr f(\bbox{r})\simeq\sum_{i,j,...}w_{i,j,...}f(\bbox{r}_{i,j,...})\,,
\end{equation}
where, unlike the case of the trapezoidal rule, the weights $w_{i,j,...}$
will not all be equal.
Special formulas for intervals near the edge can be derived,
and one can even consider variations on
the higher-order Simpson's rule.
The transverse integrations can be treated in polar coordinates
with a basis of circles of irregular radii chosen to pass through the
points of the square grid.  These methods provide results for integrals
far better than ordinary DLCQ, as shown in Fig.~\ref{fig:OneLoopSEcirc}.
A discussion of the details is
given in Appendix~\ref{sec:WeightingMethods}.

\subsubsection{Results}

The results for the momentum-space continuum limit
of the discrete sums are obtained by extrapolations
which use values of 20, 22, and 24 for $K$ and
25 through 30 for $N_\perp$.  
All results are given in units of the boson mass $\mu$.
These are fit by least
squares to either $c_0+a_1/K^3+b_1/N_\perp^2$ or
$c_0+a_1/K^3+a_2/K^4+b_1/N_\perp^2+b_2/N_\perp^3$.
The latter is used for the $\mu_1$ integral.
This means that at most 5 parameters are used in fits
to 18 points.

Extrapolation to the continuum after subtraction is
not as accurate as extrapolation of each integral
separately.  The subtraction of the discrete sums
induces a greater variation in errors that is harder to fit
properly.

The range in $N_\perp$ was selected to avoid values where
the $\mu_2$ integral was badly approximated.
However, here ``badly'' is to be interpreted relative to the
desired final error of 0.02, which is slightly more
that 0.2\% of the answer. 

\begin{figure}
\centerline{\epsfxsize=\columnwidth \epsfbox{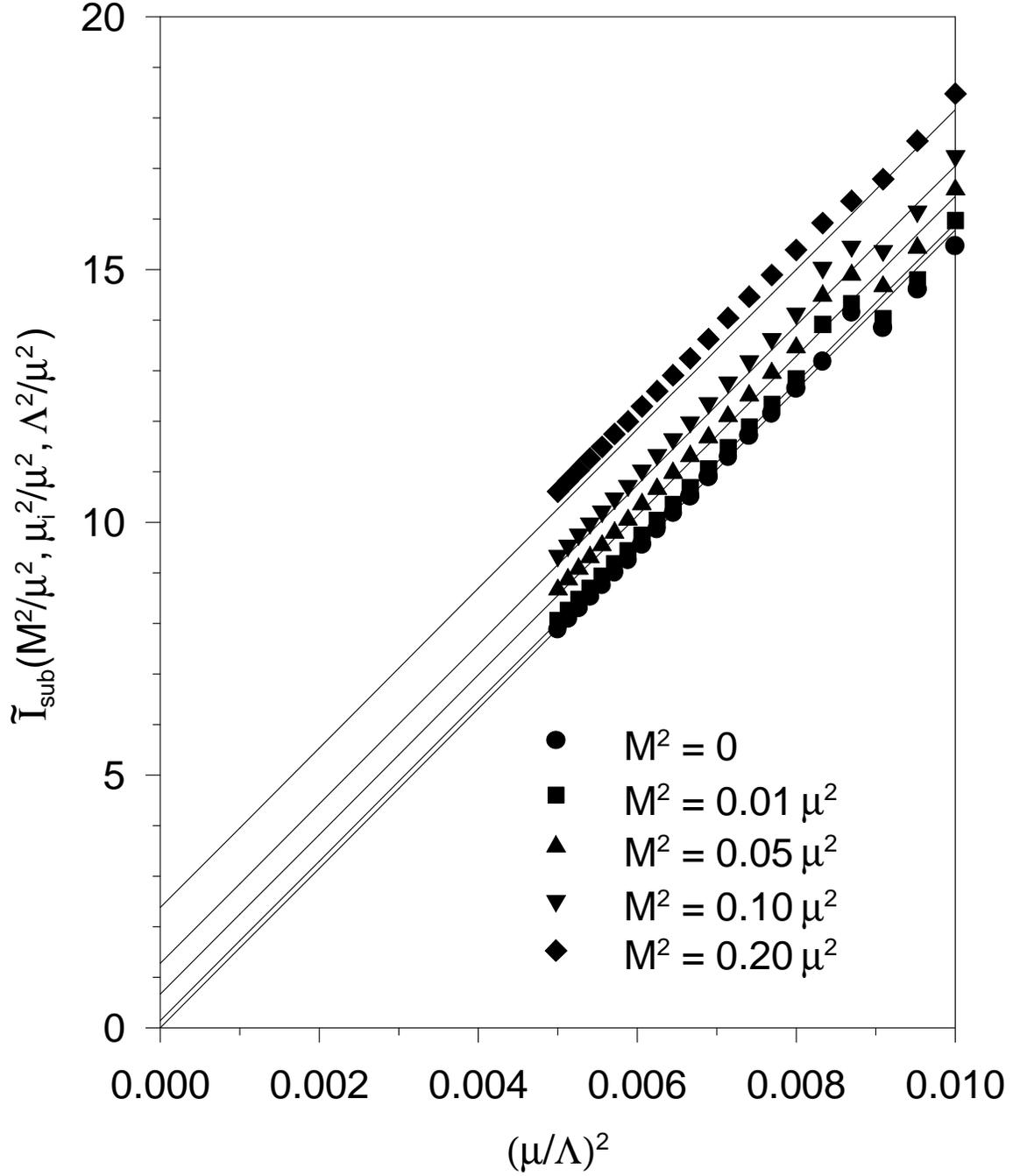} }
\caption{\label{fig:OneLoopSEsumm}
Subtracted one-loop fermion self energy.
The Pauli--Villars masses are $\mu_1^2=10\mu^2$, $\mu_2^2=50\mu^2$,
and $\mu_3^2=100\mu^2$.  The solid lines are from an analytic
expansion in $M^2$ given in (\protect\ref{eq:Iapprox}) of the text;
additional terms are needed for $M^2\protect\gtrsim 0.2\mu^2$.}
\end{figure}

Values of the subtracted integral for different
fermion masses $M$  are plotted as functions of
$1/\Lambda^2$ in Fig.~\ref{fig:OneLoopSEsumm}.  The extrapolation
to $\Lambda=\infty$ can be done by fits to 
$I_\infty+a/\Lambda^2$.  They yield the values 
in Table~\ref{tab:Linfinity}.
In obtaining these values, the error in each 
individual integral has been reduced to
$\pm0.02$ as measured against the analytic result at $M^2=0$.
This implies an error of $\pm0.04$ in the subtracted result.
Extrapolation in $\Lambda^2$ induces additional uncertainty
reflected in the miss by 0.06 of zero for $M^2=0$.
The ratios of the tabulated values are correct 
to within error estimates.  The result for the subtracted
integral is roughly proportional to $M^2$, but for $M^2$
near $0.2\mu^2$ or larger, terms even beyond $M^4$ appear important.

The range of $\Lambda^2$ values used in the fits was
from $155\mu^2$ to $200\mu^2$ in steps of $5\mu^2$.
For $\Lambda^2\leq150\mu^2$
there is some distortion.  For $\Lambda^2\leq120\mu^2$
there is significant distortion, largely due to the $\mu_3$
integral, which is badly approximated by the few points that
satisfy the cutoff.

The number of Fock states required for Pauli--Villars particles
is approximately 1.5 times the number for
physical states.  A listing of counts for two cases is
given in Table~\ref{tab:FockStates}.  Making $\mu_1$ larger does decrease
the number of Pauli--Villars states but this increases the coefficients
$\alpha_i$ and thereby amplifies errors in the integrals.  Also, with
fewer states, the integrals themselves are approximated
less accurately.

\begin{table}
\mediumtext
\caption{\label{tab:Linfinity}
Values of the subtracted integral
$\tilde{I}_{\rm sub}(M^2/\mu^2,\mu_i^2/\mu^2,\Lambda^2/\mu^2)$ in the
limit of infinite cutoff.  The Pauli--Villars masses
are $\mu_1^2=10\mu^2$, $\mu_2^2=50\mu^2$ and $\mu_3^2=100\mu^2$.}
\begin{tabular}{rddddd}
$M^2$: & 0 & 0.01$\mu^2$ & 0.05$\mu^2$ & 0.1$\mu^2$ & 0.2$\mu^2$ \\
\hline
DLCQ, improved and extrapolated: & -0.064 & 0.11 & 0.70 & 1.37 & 2.70 \\
Exact, to order $M^4$ and $M^4 \ln M^2$: & 0.0 & 0.1402 & 0.6661
                                                   & 1.2721 & 2.3778 \\
\end{tabular}
\narrowtext
\end{table}

\begin{table}
\mediumtext
\caption{\label{tab:FockStates}
Number of Fock states used in two typical cases.}
\begin{tabular}{cccccccc} 
            &     &           &          &
     \multicolumn{4}{c}{Pauli--Villars} \\ \cline{5-8}
$\Lambda^2$ & $K$ & $N_\perp$ & physical & 
     $\mu_1^2=10\mu^2$ & $\mu_2^2=50\mu^2$ & $\mu_3^2=100\mu^2$ & total\\
\hline 
200 $\mu^2$  &  20  &  25  &  25975  &  22602  & 11142 & 3305 & 37049 \\
200 $\mu^2$  &  24  &  30  &  44943  &  39162  &  19293 &  5695  &  64150 \\
\end{tabular}
\narrowtext
\end{table}

\section{A Soluble Model}

We now turn to the consideration of a nonperturbative problem.

\subsection{The effective Hamiltonian}

An effective Hamiltonian of the sort investigated by Greenberg and Schweber
\cite{SchweberGreenberg} and by G{\l}azek and Perry\cite{GlazekPerry}
can be obtained from the Yukawa Hamiltonian\cite{McCartorRobertson} by
modifying the momentum dependence in the fermion kinetic energy,
$(M^2+p_\perp^2)/p^+\longrightarrow (M_0^2+M'_0p^+)/P^+$, and by keeping only
the no-flip three-point vertex in a modified form where the longitudinal
momentum dependence is simplified.
The fermion kinetic term in the Hamiltonian has a
structure similar to that of the self-induced inertia term
shown in Eq.~(C.2) of Ref.~\cite{McCartorRobertson}.
This is a generalization of a static source.
We include one Pauli--Villars field,
which will prove sufficient in this case.  The resulting light-cone
Hamiltonian $H_{\rm LC}^{\rm eff}=P^+P_{\rm eff}^-$ is given by
\widetext
\begin{eqnarray}
\lefteqn{H_{\rm LC}^{\rm eff}=
       \int\frac{dp^+d^2p_\perp}{16\pi^3p^+}(M_0^2+M'_0p^+)
                 \sum_\sigma b_{\underline{p}\sigma}^\dagger
                                  b_{\underline{p}\sigma}} \hspace{0.5in} \\
  & +&P^+\int\frac{dq^+d^2q_\perp}{16\pi^3q^+}
       \left[\frac{\mu^2+q_\perp^2}{q^+}
                       a_{\underline{q}}^\dagger a_{\underline{q}}
           + \frac{\mu_1^2+q_\perp^2}{q^+}
                        a_{1\underline{q}}^\dagger a_{1\underline{q}}
               \right]  \nonumber \\
    & +&g\int\frac{dp_1^+d^2p_{\perp1}}{\sqrt{16\pi^3p_1^+}}
            \int\frac{dp_2^+d^2p_{\perp2}}{\sqrt{16\pi^3p_2^+}}
              \int\frac{dq^+d^2q_\perp}{16\pi^3q^+}
                \sum_\sigma b_{\underline{p}_1\sigma}^\dagger
                             b_{\underline{p}_2\sigma}
   \nonumber \\
     & &\times \left[
      \left(\frac{p_1^+}{p_2^+}\right)^\gamma
         a_{\underline{q}}^\dagger
              \delta(\underline{p}_1-\underline{p}_2+\underline{q})
        +\left(\frac{p_2^+}{p_1^+}\right)^\gamma
        a_{\underline{q}}
              \delta(\underline{p}_1-\underline{p}_2-\underline{q}) \right.
    \nonumber \\
     & & \left.
       +i\left(\frac{p_1^+}{p_2^+}\right)^\gamma
       a_{1\underline{q}}^\dagger
             \delta(\underline{p}_1-\underline{p}_2+\underline{q})
      +i\left(\frac{p_2^+}{p_1^+}\right)^\gamma
      a_{1\underline{q}}
            \delta(\underline{p}_1-\underline{p}_2-\underline{q}) \right]\,,
    \nonumber
\end{eqnarray}
with $\underline{p}\equiv(p^+,\bbox{p}_\perp)$ and
\begin{eqnarray} \label{eq:CommRelations}
\left[a_{\underline{q}},a_{\underline{q}'}^\dagger\right]
          &=&16\pi^3q^+\delta(\underline{q}-\underline{q}')\,, \\
\left\{b_{\underline{p}\sigma},b_{\underline{p}'\sigma'}^\dagger\right\}
     &=&16\pi^3p^+\delta(\underline{p}-\underline{p}')
                                      \delta_{\sigma\sigma'}\,.
\nonumber
\end{eqnarray}
The Fock-state expansion of an eigenvector is
\begin{eqnarray}
\Phi_\sigma&=&\sqrt{16\pi^3P^+}\sum_{n,n_1}
                    \int\frac{dp^+d^2p_\perp}{\sqrt{16\pi^3p^+}}
   \prod_{i=1}^n\int\frac{dq_i^+d^2q_{\perp i}}{\sqrt{16\pi^3q_i^+}}
   \prod_{j=1}^{n_1}\int\frac{dr_j^+d^2r_{\perp j}}{\sqrt{16\pi^3r_j^+}} \\
   &  & \times \delta(\underline{P}-\underline{p}
                     -\sum_i^n\underline{q}_i-\sum_j^{n_1}\underline{r}_j)
       \phi^{(n,n_1)}(\underline{q}_i,\underline{r}_j;\underline{p})
         \frac{1}{\sqrt{n!n_1!}}b_{\underline{p}\sigma}^\dagger
          \prod_i^n a_{\underline{q}_i}^\dagger 
             \prod_j^{n_1} a_{1\underline{r}_j}^\dagger |0\rangle \,.
   \nonumber
\end{eqnarray}
The normalization condition for this state is
\begin{equation}
\Phi_\sigma^{\prime\dagger}\cdot\Phi_\sigma
=16\pi^3P^+\delta(\underline{P}'-\underline{P})\,,
\end{equation}
which yields the following condition on the individual amplitudes:
\begin{equation}  \label{eq:NormCondition}
1=\sum_{n,n_1}\prod_i^n\int\,dq_i^+d^2q_{\perp i}
                     \prod_j^{n_1}\int\,dr_j^+d^2r_{\perp j}
    \left|\phi^{(n,n_1)}(\underline{q}_i,\underline{r}_j;
           \underline{P}-\sum_i\underline{q}_i
                              -\sum_j\underline{r}_j)\right|^2\,.
\end{equation}

\subsection{Analytic solution}

We seek a solution to
\begin{equation}
H_{\rm LC}^{\rm eff}\Phi_\sigma=M^2\Phi_\sigma\,.
\end{equation}
With $y_i=q_i^+/P^+$ and $z_j=r_j^+/P^+$, the amplitudes must then satisfy
\begin{eqnarray} \label{eq:CoupledEqns}
\lefteqn{\left[M^2-M_0^2-M'_0p^+
  -\sum_i\frac{\mu^2+q_{\perp i}^2}{y_i}
                  -\sum_j\frac{\mu_1^2+r_{\perp j}^2}{z_j}\right]
                    \phi^{(n,n_1)}(\underline{q}_i,
                           \underline{r}_j,\underline{p})} \hspace{0.2in} \\
& =g&\left\{\sqrt{n+1}\int\frac{dq^+d^2q_\perp}{\sqrt{16\pi^3q^+}}
              \left(\frac{p^+-q^+}{p^+}\right)^\gamma
              \phi^{(n+1,n_1)}(\underline{q}_i,\underline{q},
                    \underline{r}_j,\underline{p}-\underline{q})\right.
\nonumber \\
& & +\frac{1}{\sqrt{n}}\sum_i\frac{1}{\sqrt{16\pi^3q_i^+}}
              \left(\frac{p^+}{p^++q_i^+}\right)^\gamma
              \phi^{(n-1,n_1)}(\underline{q}_1,\ldots,\underline{q}_{i-1},
                      \underline{q}_{i+1},\ldots,\underline{q}_n,
                       \underline{r}_j,\underline{p}+\underline{q}_i)
\nonumber \\
& &+i\sqrt{n_1+1}\int\frac{dr^+d^2r_\perp}{\sqrt{16\pi^3r^+}}
              \left(\frac{p^+-r^+}{r^+}\right)^\gamma
              \phi^{(n,n_1+1)}(\underline{q}_i,\underline{r}_j,
                           \underline{r},\underline{p}-\underline{r})
\nonumber \\
& & +\left.\frac{i}{\sqrt{n_1}}\sum_j\frac{1}{\sqrt{16\pi^3r_j^+}}
              \left(\frac{p^+}{p^++r_j^+}\right)^\gamma
              \phi^{(n,n_1-1)}(\underline{q}_i,\underline{r}_1,\ldots,
                                     \underline{r}_{j-1},
                        \underline{r}_{j+1},\ldots,\underline{r}_{n_1},
                           \underline{p}+\underline{r}_j) \right\}\,.
\nonumber
\end{eqnarray}
By construction, this coupled set of integral equations is identical
in basic form to the equations considered by Greenberg and Schweber
\cite{SchweberGreenberg}.  Their factorized {\em ansatz} for a
solution suggests that we try
\begin{equation}   \label{eq:AnalyticSoln}
\phi^{(n,n_1)}=\sqrt{Z}\frac{(-g)^n(-ig)^{n_1}}{\sqrt{n!n_1!}}
         \left(\frac{p^+}{P^+}\right)^\gamma
         \prod_i\frac{y_i}{\sqrt{16\pi^3q_i^+}(\mu^2+q_{\perp i}^2)}
          \prod_j\frac{z_j}{\sqrt{16\pi^3r_j^+}(\mu_1^2+r_{\perp j}^2)}\,.
\end{equation}
\narrowtext
\noindent
This is indeed a solution, provided that $M_0=M$ and
\begin{equation}
M'_0=\frac{g^2/P^+}{16\pi^2}\frac{\ln\mu_1/\mu}{\gamma+1/2}\,.
\end{equation}
Although $\gamma$ can be assigned any of a range of values, $1/2$ is the
natural choice, and we will use this value for the remainder of the paper.
With this choice, the one-boson amplitude is proportional to $\sqrt{y(1-y)}$.

The normalization condition (\ref{eq:NormCondition}) implies
\begin{equation}
\frac{1}{Z}=\sum_{n,n_1} \frac{1}{(2n+2n_1+1)!n!n_1!}
              \frac{(g/\mu)^{2n} (g/\mu_1)^{2n_1}}{(16\pi^2)^{n+n_1}}\,.
\end{equation}
Thus we can fix the bare mass and the wave function renormalization.  However,
there remains the bare coupling.\footnote{In this model the bare coupling
is finite.}

\subsection{Coupling renormalization}

To fix the coupling we use $\langle :\!\!\phi^2(0)\!\!:\rangle
\equiv\Phi_\sigma^\dagger\!:\!\!\phi^2(0)\!\!:\!\Phi_\sigma$.  For the
analytic solution this expectation value reduces to
\begin{equation}
\langle :\!\!\phi^2(0)\!\!:\rangle=
      \sum_{n,n_1} \frac{2Zn}{(2n+2n_1)!n!n_1!}
      \frac{(g/\mu)^{2n} (g/\mu_1)^{2n_1}}{(16\pi^2)^{n+n_1}}  \,.
\end{equation}
From a numerical solution it can be computed fairly efficiently in a sum
similar to the normalization sum
\widetext
\begin{eqnarray}
\langle :\!\!\phi^2(0)\!\!:\rangle
        =&\sum_{n=1,n_1=0}\prod_i^n&\int\,dq_i^+d^2q_{\perp i}
                     \prod_j^{n_1}\int\,dr_j^+d^2r_{\perp j}
                     \left(\sum_{k=1}^n \frac{2}{q_k^+/P^+}\right) \\
    & &\times \left|\phi^{(n,n_1)}(\underline{q}_i,\underline{r}_j;
       \underline{P}-\sum_i\underline{q}_i-\sum_j\underline{r}_j)\right|^2\,.
    \nonumber
\end{eqnarray}

With the bare parameters determined, we ``predict'' a value for the
slope of the fermion no-flip form factor.  It is related to the
transverse size of the dressed fermion.
From\cite{BrodskyDrell} we find a useful expression for the
form factor
\begin{eqnarray}
F(Q^2)&=&\frac{1}{2P^+}\langle P+p_\gamma\uparrow |J^+(0)|P\uparrow\rangle \\
      &=&\sum_j e_j\int 16\pi^3\delta(1-\sum_i x_i)
                   \delta(\sum_i \bbox{k}_{\perp i})
      \prod_i \frac{dx_id^2p_{\perp i}}{16\pi^3}
      \psi_{P+p_\gamma\uparrow}^*(x_i,\bbox{p}'_{\perp i})
              \psi_{P\uparrow}(x_i,\bbox{p}_{\perp i})\,,    \nonumber
\end{eqnarray}
where the matrix element has been evaluated in the frame with
\begin{equation}
P=(P^+,P^-=\frac{M^2}{P^+},\bbox{0}_\perp)\,,\;\;
p_\gamma=(0,p_\gamma^-=2p_\gamma\cdot P/P^+,\bbox{p}_{\gamma\perp})\,,\;\;
 Q^2\equiv p_{\gamma\perp}^2\,,
\end{equation}
$e_j$ is the charge of the jth constituent, and
\begin{equation}
\bbox{p}'_{\perp i}=\left\{\begin{array}{cc}
             \bbox{p}_{\perp i}-x_i\bbox{p}_{\gamma\perp} & i\neq j \\
             \bbox{p}_{\perp i}+(1-x_i)\bbox{p}_{\gamma\perp} & i=j
             \,. \end{array}\right.
\end{equation}
A sum over Fock states is understood.

When the fermion is assigned a charge of 1, and the bosons remain neutral,
the analytic solution yields
\begin{eqnarray}
F(Q^2)=&Z\sum_{n,n_1} &\frac{(g^2/16\pi^3)^{n+n_1}}{n!n_1!}
    \int_0^1\theta(1-\sum_i^n y_i-\sum_j^{n_1}z_j) \\
&&  \times
      \prod_i^n \frac{y_idy_id^2q_{\perp i}}
                      {(\mu^2+q_{\perp i}^{\prime 2})(\mu^2+q_{\perp i}^2)}
      \prod_j^{n_1} \frac{z_j dz_jd^2r_{\perp j}}
                      {(\mu^2+r_{\perp j}^{\prime 2})(\mu^2+r_{\perp j}^2)}
                \,, \nonumber
\end{eqnarray}
with  
\begin{equation}
\bbox{q}'_\perp=\bbox{q}_\perp-y\bbox{p}_{\gamma\perp}\;\; \mbox{and}\;\;
\bbox{r}'_\perp=\bbox{r}_\perp-z\bbox{p}_{\gamma\perp}\,.
\end{equation}
The slope is extracted as
\begin{equation}
F'(0)=-\sum_{n,n_1} \frac{Z(n/\mu^2+n_1/\mu_1^2)}{(2n+2n_1+3)!n!n_1!}
          \frac{(g/\mu)^{2n} (g/\mu_1)^{2n_1}}{(16\pi^2)^{n+n_1}}\,.
\end{equation}
Numerically, one can compute $F'(0)$ from
\begin{eqnarray} \label{eq:Fprime}
\lefteqn{F'(0)=\sum_{n,n_1}\prod_i^n\int\,dq_i^+d^2q_{\perp i}
                   \prod_j^{n_1}\int\,dr_j^+d^2r_{\perp j} }\hspace{0.5in}\\
     & \times \left[\left(\sum_i \frac{y_i^2}{4}
                             \nabla_{\perp i}^2\right.\right. &
         +\left.\left. \sum_j \frac{z_j^2}{4}\nabla_{\perp j}^2\right)
                   \phi^{(n,n_1)}(\underline{q}_i,\underline{r}_j;
       \underline{P}-\sum_i\underline{q}_i-\sum_j\underline{r}_j)\right]^*
    \nonumber \\
    & &  \times
             \phi^{(n,n_1)}(\underline{q}_i,\underline{r}_j;
           \underline{P}-\sum_i\underline{q}_i-\sum_j\underline{r}_j)\,,
    \nonumber
\end{eqnarray}
with $\nabla_\perp^2$ represented by finite differences.
It turns out that single derivatives of typical amplitudes can
be better approximated than the double derivatives in the
Laplacian.  Integration by parts in (\ref{eq:Fprime}) then leads
to a computationally better quantity
\begin{eqnarray} \label{eq:BetterFprime}
\lefteqn{\tilde{F}'(0)=-\sum_{n,n_1}\prod_i^n\int\,dq_i^+d^2q_{\perp i}
                   \prod_j^{n_1}\int\,dr_j^+d^2r_{\perp j}}\hspace{0.5in} \\
     & \times & \left[\sum_i \left|\frac{y_i}{2}\nabla_{\perp i}
             \phi^{(n,n_1)}(\underline{q}_i,\underline{r}_j;
           \underline{P}-\sum_i\underline{q}_i-\sum_j\underline{r}_j)
                                        \right|^2 \right.
    \nonumber \\
    & & \left. +\sum_j \left|\frac{z_j}{2}\nabla_{\perp j}
                   \phi^{(n,n_1)}(\underline{q}_i,\underline{r}_j;
           \underline{P}-\sum_i\underline{q}_i
              -\sum_j\underline{r}_j)\right|^2 \right]\,,
    \nonumber
\end{eqnarray}
which differs from $F'(0)$ by surface terms which vanish as
$\Lambda\rightarrow\infty$.

\subsection{Distribution functions}

To further explore the wave functions $\phi^{(n,n_1)}$, we compute
distribution functions for the constituent bosons
\begin{eqnarray}
f_B(y)\equiv&\sum_{n,n_1}\prod_i^n\int\,dq_i^+d^2q_{\perp i}
                     \prod_j^{n_1}\int&\,dr_j^+d^2r_{\perp j}
       \sum_{i=1}^n\delta(y-q_i^+/P^+) \\
   &&\times \left|\phi^{(n,n_1)}(\underline{q}_i,\underline{r}_j;
           \underline{P}-\sum_i\underline{q}_i
                             -\sum_j\underline{r}_j)\right|^2\,,
   \nonumber
\end{eqnarray}
and the Pauli--Villars boson
\begin{eqnarray}
f_{PV}(z)\equiv&\sum_{n,n_1}\prod_i^n\int\,dq_i^+d^2q_{\perp i}
                     \prod_j^{n_1}\int&\,dr_j^+d^2r_{\perp j}
       \sum_{j=1}^{n_1}\delta(z-r_j^+/P^+)  \\
  &&\times  \left|\phi^{(n,n_1)}(\underline{q}_i,\underline{r}_j;
           \underline{P}-\sum_i\underline{q}_i
                          -\sum_j\underline{r}_j)\right|^2\,.
   \nonumber
\end{eqnarray}
Their integrals yield the average multiplicities
\begin{equation}
\langle n_B\rangle=\int_0^1f_B(y)dy\,,\;\;
\langle n_{PV}\rangle=\int_0^1f_{PV}(z)dz\,.
\end{equation}
For the analytic solution (\ref{eq:AnalyticSoln}) we obtain
\begin{equation}
f_B(y)=\left(\frac{\mu_1}{\mu}\right)^2f_{PV}(y)
      =\sum_{n,n_1} \frac{Zny(1-y)^{(2n+2n_1-1)}}{(2n+2n_1-1)!n!n_1!}
                \frac{(g/\mu)^{2n} (g/\mu_1)^{2n_1}}{(16\pi^2)^{n+n_1}}\,,
\end{equation}
and
\begin{equation}
\langle n_B\rangle=\left(\frac{\mu_1}{\mu}\right)^2\langle n_{PV}\rangle
    =\sum_{n,n_1} \frac{Zn}{(2n+2n_1+1)!n!n_1!}
                  \frac{(g/\mu)^{2n} (g/\mu_1)^{2n_1}}{(16\pi^2)^{n+n_1}}\,.
\end{equation}
For a numerical solution, the integrals can be approximated by sums.

\narrowtext

\section{DLCQ Applied to the Soluble Model} \label{sec:DLCQapplied}

\subsection{Discretization}

The basic momentum discretization and approximation of integrals
are discussed in Sec.~\ref{sec:DLCQ}\@.  From these we construct
discrete approximations to the eigenvector $\Phi_\sigma$,
the coupled equations (\ref{eq:CoupledEqns}) for the amplitudes,
and the derived quantities $\langle:\!\!\phi^2(0)\!\!:\rangle$,
$\tilde{F}'$ and distribution amplitudes.  Creation
operators for discrete momenta are defined by
\begin{equation}
b_{\underline{n}\sigma}^\dagger=
    \frac{\pi/L_\perp}{\sqrt{8\pi^3 n}}b_{\underline{p}\sigma}^\dagger\,,\;\;
a_{\underline{m}}^\dagger=
    \frac{\pi/L_\perp}{\sqrt{8\pi^3 m}}a_{\underline{q}}^\dagger\,,\;\;
\end{equation}
such that they satisfy simple commutation relations
\begin{equation}
\left\{b_{\underline{n}\sigma},b_{\underline{n}\sigma}^\dagger\right\}=
  \delta_{\underline{n}'\underline{n}}\delta_{\sigma'\sigma}\,,\;\;
\left[a_{\underline{n}},a_{\underline{n}}^\dagger\right]=
  \delta_{\underline{n}'\underline{n}}\,.
\end{equation}
These follow from (\ref{eq:CommRelations}) and the discrete
delta-function representation
\begin{equation}
\delta(\underline{p}-\underline{p}')=
  \frac{L}{2\pi}\left(\frac{L_\perp}{\pi}\right)^2
               \delta_{\underline{n}'\underline{n}}\,.
\end{equation}
The discrete approximation of the eigenvectors is then
\begin{eqnarray}
\tilde{\Phi}_\sigma\equiv\frac{\pi}{L_\perp}\Phi_\sigma
   &=&\sqrt{8\pi^3 K}\sum_{n,n_1}\sum_{\underline{n}}
        \prod_{i=1}^n \sum_{\underline{m}_i}
   \prod_{j=1}^{n_1}\sum_{\underline{l}_j}
   \delta_{\underline{K},\underline{n}
                     +\sum_i^n\underline{m}_i+\sum_j^{n_1}\underline{l}_j}
                     \nonumber  \\
    &  & \times  \tilde{\phi}^{(n,n_1)}(\underline{m}_i,\underline{l}_j;
                                                             \underline{n})
         \frac{1}{\sqrt{n!n_1!}}b_{\underline{n}\sigma}^\dagger
          \prod_i^n a_{\underline{m}_i}^\dagger
             \prod_j^{n_1} a_{1\underline{l}_j}^\dagger |0\rangle \,,
   \nonumber
\end{eqnarray}
where
\begin{equation}
\tilde{\phi}^{(n,n_1)}=\left[\frac{2\pi}{L}
      \left(\frac{\pi}{L_\perp}\right)^2\right]^{(n+n_1)/2}\phi^{(n,n_1)}
\end{equation}
are rescaled amplitudes, for which the normalization condition
(\ref{eq:NormCondition}) becomes
\begin{equation}
1=\sum_{n,n_1} \prod_{i=1}^n \sum_{\underline{m}_i}
   \prod_{j=1}^{n_1}\sum_{\underline{l}_j}
   \left|\tilde{\phi}^{(n,n_1)}(\underline{m}_i,\underline{l}_j,
         \underline{K}-\sum_i\underline{m}_i-\sum_j\underline{l}_j)
         \right|^2\,.
\end{equation}

The most convenient basis for a numerical calculation  is the number
basis (or oscillator basis), which eliminates summation over states
that differ by only rearrangement of bosons of the same type.  We
define collections of sums with a prime
$\prod_{i=1}^n \sum_{\underline{m}_i}^\prime$ as being restricted
to one ordering of the momenta and introduce factorials
$N_{\{\underline{m}_i\}}\equiv
N_{\underline{m}_1}!N_{\underline{m}_2}!\cdots$ where
$N_{\underline{m}_1}$ is the number of times that
$\underline{m}_1$ appears in the collection
$\{\underline{m}_i\}$.  The amplitudes for this number basis
are
\begin{equation}
\psi^{(n,n_1)}=\sqrt{\frac{n!n_1!}
                {N_{\{\underline{m}_i\}}N_{\{\underline{l}_j\}}}}
                \tilde{\phi}^{(n,n_1)}\,,
\end{equation}
with normalization
\begin{equation}
1=\sum_{n,n_1} \prod_{i=1}^n \sum_{\underline{m}_i}\,^\prime
   \prod_{j=1}^{n_1}\sum_{\underline{l}_j}\,^\prime
   \left|\psi^{(n,n_1)}\right|^2\,.
\end{equation}
In this basis the discretization of the coupled equations
(\ref{eq:CoupledEqns}) yields
\widetext
\begin{eqnarray}   \label{eq:MatrixEq}
\lefteqn{\left[\tilde{M}^2-\tilde{M}_0^2-\tilde{M}'_0\frac{n}{K}
    -\sum_i\frac{1+(m_{ix}^2+m_{iy}^2)/\tilde{L}_\perp^2}{m_i/K}
  -\sum_j\frac{\tilde{\mu}_1^2+(l_{jx}^2+l_{jy}^2)/\tilde{L}_\perp^2}{l_j/K}
       \right]\psi^{(n,n_1)}(\underline{m}_i,\underline{l}_j,\underline{n})}
      \hspace{0.2in} \nonumber \\
&  =&\frac{g/\mu}{\tilde{L}_\perp\sqrt{8\pi^3}}
   \left\{\sum_{\underline{m}}
        \frac{1}{\sqrt{m}}
        \sqrt{\frac{N_{\{\underline{m},\underline{m}_i\}} }
                     {N_{\{\underline{m}_i\}} }}
        \left(\frac{n-m}{n}\right)^\gamma
              \psi^{(n+1,n_1)}(\underline{m}_i,\underline{m},
                    \underline{l}_j,\underline{n}-\underline{m})\right.
 \\
& &  +\sum_i\frac{1}{\sqrt{m_i}}
        \sqrt{\frac{N_{\{\underline{m}_i\}'}    }
                     {N_{\{\underline{m}_i\}} }}
        \left(\frac{n}{n+m_i}\right)^\gamma
              \psi^{(n-1,n_1)}(\underline{m}_1,\ldots,\underline{m}_{i-1},
                        \underline{m}_{i+1},\ldots,\underline{m}_n,
                          \underline{l}_j,\underline{n}+\underline{m}_i)
\nonumber \\
& &+i\sum_{\underline{l}}
        \frac{1}{\sqrt{l}}
        \sqrt{\frac{N_{\{\underline{l},\underline{l}_j\}} }
                     {N_{\{\underline{l}_j\}} }}
        \left(\frac{n-l}{n}\right)^\gamma
              \psi^{(n,n_1+1)}(\underline{m}_i,\underline{l}_j,
                      \underline{l},\underline{n}-\underline{l})
\nonumber \\
& & +\left.i\sum_j\frac{1}{\sqrt{l_j}}
        \sqrt{\frac{N_{\{\underline{l}_j\}'}}
                     {N_{\{\underline{l}_j\}} }}
        \left(\frac{n}{n+l_j}\right)^\gamma
              \psi^{(n,n_1-1)}(\underline{m}_i,\underline{l}_1,\ldots,
                      \underline{l}_{j-1},\underline{l}_{j+1},\ldots,
              \underline{l}_{n_1},\underline{n}+\underline{l}_j) \right\}\,,
\nonumber
\end{eqnarray}
\narrowtext%
\noindent where
$\underline{n}=\underline{K}-\sum_i\underline{m}_i-\sum_j\underline{l}_j$,
$\{\underline{m}_i\}'$ is the
set of boson momenta without $\underline{m}_i$, and a tilde implies
division by $\mu$ except for $\tilde{L}_\perp=\mu L_\perp/\pi$.
This is a matrix eigenvalue problem, which for given $g$, $\mu$,
$M$, $\mu_1$, and $\Lambda$ we solve for $\psi$ and $M_0^2$.
The cutoff $\Lambda^2$ is applied as a limit on the invariant mass of
individual particles, rather than on the total invariant mass of a
Fock state.  Typical basis sizes are given in Table~\ref{tab:basis50}.

\begin{table}
\mediumtext
\caption{\label{tab:basis50}
Basis sizes for DLCQ calculations in the soluble model
with parameters $M^2=\mu^2$, $\mu_1^2=10\mu^2$, and 
$\Lambda^2=50\mu^2$.  The numbers of physical states
are in parentheses.}
\begin{tabular}{c|rrrrrrrrrr}
  & \multicolumn{8}{c}{K} \\
 \cline{2-9}
$N_\perp$ & 3 & 5 &  7   &   9    &     11  &    13   &     15  & 17    \\
\hline
1  &   3 &     8 &    18 &     38 &      36 &      65 &     110 &    185 \\
   &(  2)&(    4)&(    7)& (   12)&  (   19)&  (   30)&  (   45)& (   67)\\
2  &  19 &    70 &   218 &    265 &     590 &    1120 &     822 &   1410 \\
   &( 10)&(   32)&(  127)& (  119)&  (  343)&  (  754)&  (  453)& (  626)\\
3  &  43 &   222 &   958 &   1408 &    4460 &   17031 &   22486 &  21635 \\
   &( 22)&(  102)&(  367)& (  736)&  ( 2671)&  ( 9230)&  (13213)& (13531)\\
4  &  75 &   872 &  3714 &   9259 &   49394 &   50966 &  110254 & 328966 \\
   &( 38)&(  330)&( 1399)& ( 5913)&  (32363)&  (32124)&  (55319)&(172247)\\
5  &  99 &  2028 & 13702 &  54100 &   95176 &  386140 & 1553576  \\
   &( 50)&(  722)&( 5699)& (28065)&  (66371)& (232400)&(1038070)  \\
6  & 139 &  3982 & 35666 & 126748 &  536758 & 2907158  \\
   &( 70)&( 1548)&(12991)& (69245)& (391511)&(2107688)  \\
7  & 195 &  7734 & 79794 & 519325 & 1317392  \\
   &( 98)&( 2780)&(32891)&(276299)&(1008539)  \\
8  & 275 & 11736 &172118 &1165832  \\
   &(138)&( 4268)&(61947)&(687394)  \\
\end{tabular}
\narrowtext
\end{table}

The bare coupling $g$ is fixed by setting
a value for
\begin{equation} \label{eq:phiSq}
\langle:\!\!\phi^2(0)\!\!:\rangle\simeq\sum_{n,n_1}
   \prod_{i=1}^n \sum_{\underline{m}_i}\,^\prime
     \prod_{j=1}^{n_1}\sum_{\underline{l}_j}\,^\prime
        \sum_{k=1}^n\frac{2K}{m_k} \left|\psi^{(n,n_1)}\right|^2\,.
\end{equation}
The combination of the matrix equation and the imposed constraint on
$\langle:\!\!\phi^2(0)\!\!:\rangle$ is solved iteratively.  The form factor
slope $\tilde{F}'(0)$, the distribution functions, and average
multiplicities are all approximated by similar discrete sums over the
amplitudes $\psi^{(n,n_1)}$.

\subsection{Numerical techniques}

The matrix equation (\ref{eq:MatrixEq}) is solved using the
Lanczos algorithm \cite{Lanczos} for complex symmetric
matrices, which is a special case of the biorthogonal
Lanczos algorithm \cite{Biorthogonal,Cullum}.
Given an initial guess $\bbox{u}_1$
for an eigenvector of a complex symmetric matrix $A$,
a sequence of vectors $\{\bbox{u}_n\}$ is generated by
the following steps:
\begin{eqnarray}
\bbox{v}_{n+1}&=&A\bbox{u}_n-b_n\bbox{u}_{n-1}\;\;\;
(\mbox{with}\;b_1=0) \nonumber \\
a_n&=&\bbox{v}_{n+1}\cdot\bbox{u}_n \nonumber \\
\bbox{v}_{n+1}^\prime&=&\bbox{v}_{n+1}-a_n\bbox{u}_n \\
b_{n+1}&=&\sqrt{\bbox{v}_{n+1}^\prime\cdot\bbox{v}_{n+1}^\prime}
\nonumber \\
\bbox{u}_{n+1}&=&\bbox{v}_{n+1}^\prime/b_{n+1}\,.   \nonumber
\end{eqnarray}
The dot products do not involve conjugation, and the constants $a_n$ and
$b_n$ are in general complex.  The process will fail if
$b_{n+1}$ is zero for nonzero $\bbox{v}_{n+1}^\prime$, which can
happen in principle but does not seem to happen in practice \cite{Cullum}.
If $\bbox{v}_{n+1}^\prime$ is zero, the process terminates naturally.
The vectors $\bbox{u}_{n-1}$, $\bbox{v}_{n+1}$, $\bbox{v}_{n+1}^\prime$,
and $\bbox{u}_{n+1}$ can all be stored in the same array.  At any
one time only two vectors, one of these and $\bbox{u}_n$, need to be kept.

The vectors $\bbox{u}_n$ are orthogonal to each other, and the $a_n$ and
$b_n$ form the diagonal and co-diagonal of a complex symmetric tridiagonal
matrix which represents $A$ in the basis $\{\bbox{u}_n\}$.  If the process
has terminated with a $\bbox{v}_{n+1}^\prime=0$ for some $n$, the tridiagonal
representation is an exact representation for some subspace, and
diagonalization yields some of the eigenvalues of $A$.  If the process
is terminated at some arbitrary early point, the eigenvalues of the
tridiagonal matrix will approximate those of $A$.  The approximation
is particularly good for the extreme eigenvalues after only a few
iterations.  Depending on the initial guess, the number of iterations
may need to be only 20, independent of the size of $A$.  To reconstruct
the eigenvectors of the original matrix, all of the $\bbox{u}_n$ need
to be kept.  Because only two are needed in the Lanczos algorithm,
the others can be written temporarily to disk and retrieved later.

We use the analytic solution (\ref{eq:AnalyticSoln}) as the initial
guess.  Its components are either real or imaginary, and the process of
matrix multiplication and division/multiplication by $b_n$ preserves
this structure in a controlled way.  The diagonal elements $a_n$ can
be shown to be real and the off-diagonal elements $b_n$ are either
real or imaginary.  This reduces the storage needed and eliminates
the need for explicit complex arithmetic.

To further reduce storage requirements, we take full advantage of the
transposition symmetry and sparsity of the matrix.  Only nonzero elements
and their indices are stored.  The coupling $g$ is factored out so that
the matrix can be reused without change in the iterations that solve
for $g$.

To improve convergence, we include weighting factors of the sort
discussed in Sec.~\ref{sec:DLCQ} and Appendix~\ref{sec:WeightingMethods}.
The circular form for transverse sums is used for two-body
amplitudes and the extended trapezoidal rule is used for all others,
with one exception.  If the coefficients (\ref{eq:TrapCoefficients})
for the extended trapezoidal rule become negative, a rectangle approximation
is used.  This is because restoration of symmetry for the weighted
matrix requires the square roots of the weights.  Schematically
the symmetrization process is
\begin{equation}
\sum_j A_{ij}w_ju_j=\xi u_i\;\;\longrightarrow\;\;
\sum_j\sqrt{w_i w_j}A_{ij} \sqrt{w_j}u_j=\xi\sqrt{w_i} u_i\,,
\end{equation}
where $\sqrt{w_i w_j}A_{ij}$ is the new symmetric matrix.

The complete specification of the weighting factors requires
selection of integration order, because the limits of integration
are interrelated by the cutoff.  The simplest reduction of these
interrelationships is made if all sums in one transverse direction are done
before those in the orthogonal transverse direction, and all of
these before the longitudinal sums.  Within each of these three
groupings the sums are done for one particle at a time in the
ordered momentum list.  One consequence of this choice is that
the transverse directions are treated asymmetrically
(except for the two-body sectors).  This induces a small transverse
asymmetry in the amplitudes of the solution, including the
two-body amplitudes.  The asymmetry disappears in the numerical
limit $N_\perp\rightarrow\infty$.

\subsection{Results}

We have solved the discrete eigenvalue problem (\ref{eq:MatrixEq})
for various cases.  The physical parameter values chosen
were $\gamma=1/2$, $M^2=\mu^2$, $\mu_1^2=10\mu^2$, and
$\langle :\!\!\phi^2(0)\!\!:\rangle=1$ or 2.  The parameters
that control the numerical approximation, namely $\Lambda^2$,
$K$, and $N_\perp$ (or $L_\perp$), were varied to study
convergence with basis sizes up to $\sim520,000$.  The ranges
of these numerical parameters are shown in Tables~\ref{tab:phi1}
and \ref{tab:phi2}.  The transverse scale $L_\perp$ was chosen
such that $N_\perp$ radial points satisfy the invariant-mass cutoff
for one-boson states at the value of longitudinal momentum that
yields maximum transverse width.  The bare fermion mass $M_0$ was
allowed to vary from its analytic, infinite-cutoff value of $M$
in order that $M$ could be held fixed.  The tables list the
values of $M_0$ along with those of the bare coupling $g$,
as set by (\ref{eq:phiSq}), the average boson multiplicity
$\langle n_B\rangle$, and, for $\langle:\!\!\phi^2(0)\!\!:\rangle=1$,
the slope $\tilde{F}'(0)$ of the fermion form factor.  The analytic,
infinite-cutoff values are also included.

\begin{table}
\caption{\label{tab:phi1}
Numerical parameter values and results from solving 
the model eigenvalue problem.  The physical parameter
values were $M^2=\mu^2$ for the fermion mass, $\mu_1^2=10\mu^2$
for the Pauli--Villars mass, and $\langle:\!\!\phi^2(0)\!\!:\rangle=1$
to fix the coupling $g$.}
\begin{tabular}{cccccccc}
$(\Lambda/\mu)^2$ & $K$ & $N_\perp$ & $\mu L_\perp/\pi$ & 
  $(M_0/\mu)^2$ & $g/\mu$ & $\langle n_B\rangle$ & $100\mu^2\tilde{F}'(0)$ \\
\hline
 50 & 11 & 4 & 0.8165 & 0.8547 & 13.293 & 0.177 & -0.751 \\
 50 & 13 & 4 & 0.8165 & 0.8518 & 13.230 & 0.172 & -1.015 \\
 50 & 15 & 4 & 0.8165 & 0.8408 & 13.556 & 0.178 & -0.715 \\
 50 & 17 & 4 & 0.8165 & 0.8289 & 13.392 & 0.180 & -0.565 \\
 \\
 50 &  9 & 5 & 1.2062 & 0.8601 & 14.023 & 0.179 & -0.547 \\
 50 &  9 & 6 & 1.2247 & 0.8377 & 14.323 & 0.179 & -0.582 \\
 50 &  9 & 7 & 1.4289 & 0.8302 & 14.386 & 0.179 & -0.658 \\
 \\
 50 &  9 & 5 & 1.2062 & 0.8601 & 14.023 & 0.179 & -0.547 \\
100 &  9 & 5 & 0.7143 & 1.0520 & 12.565 & 0.174 & -0.239 \\
200 &  9 & 5 & 0.5025 & 1.1980 & 10.191 & 0.172 & -0.139 \\
 \\
$\infty$  &  \multicolumn{3}{c}{analytic}
                 & 1.0000 & 13.148 & 0.160 & -0.786 \\
\end{tabular}
\end{table}

\begin{table}
\caption{\label{tab:phi2}
Same as Table~\protect\ref{tab:phi1} except
$\langle:\!\!\phi^2(0)\!\!:\rangle=2$.}
\begin{tabular}{ccccccc}
$(\Lambda/\mu)^2$ & $K$ & $N_\perp$ & $\mu L_\perp/\pi$ & 
  $(M_0/\mu)^2$ & $g/\mu$ & $\langle n_B\rangle$ \\
\hline
 50 & 11 & 4 & 0.8165 & 0.5068 & 21.541 & 0.368  \\
 50 & 13 & 4 & 0.8165 & 0.5166 & 21.327 & 0.352  \\
 50 & 15 & 4 & 0.8165 & 0.4496 & 22.323 & 0.366  \\
 50 & 17 & 4 & 0.8165 & 0.4439 & 21.930 & 0.364  \\
 \\
 50 &  9 & 5 & 1.2062 & 0.5340 & 22.396 & 0.367  \\
 50 &  9 & 6 & 1.2247 & 0.5109 & 22.507 & 0.369  \\
 50 &  9 & 7 & 1.4289 & 0.5204 & 22.287 & 0.366  \\
 \\
 50 &  9 & 5 & 1.2062 & 0.5340 & 22.396 & 0.367   \\
100 &  9 & 5 & 0.7143 & 0.9353 & 20.962 & 0.359   \\
200 &  9 & 5 & 0.5025 & 1.3080 & 18.034 & 0.347  \\
 \\
$\infty$  &  \multicolumn{3}{c}{analytic} & 1.0000 & 19.420 & 0.308 \\
\end{tabular}
\end{table}

The values of the form factor slope are very poor
approximations.  This is due to the sensitivity to $N_\perp$ of
the finite difference representation of the derivatives
in (\ref{eq:BetterFprime}).  A good approximation requires at least
$N_\perp=8$, which implies very large basis sizes even for
small $K$.

\begin{figure}
\centerline{\epsfxsize=\columnwidth \epsfbox{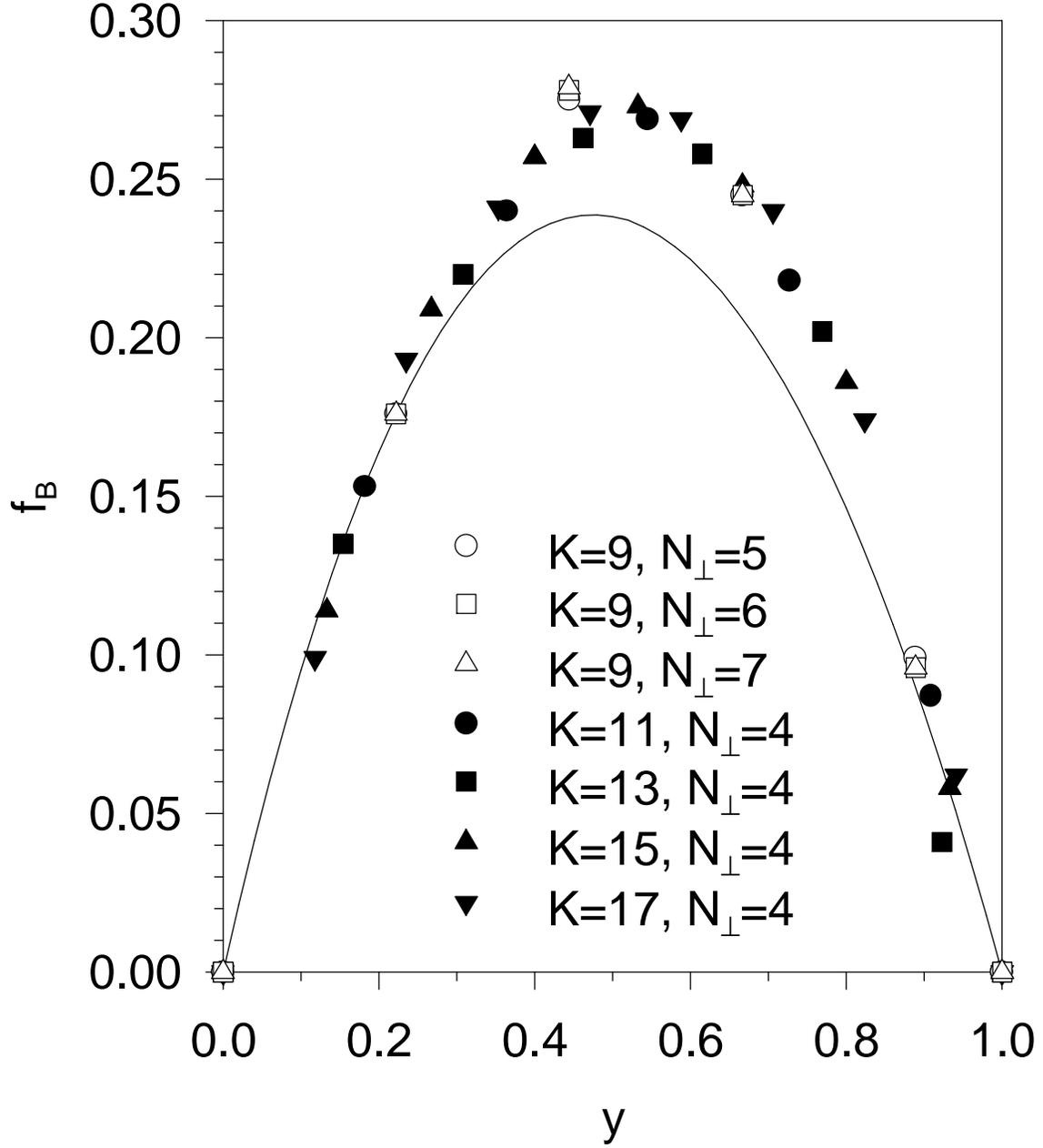} }
\caption{\label{fig:fBNK-phi1}
The boson distribution function $f_B$ at various numerical resolutions,
with $\langle:\!\!\phi^2(0)\!\!:\rangle=1$
and $\Lambda^2=50\mu^2$.  The solid line is the analytic result.}
\end{figure}

\begin{figure}
\centerline{\epsfxsize=\columnwidth \epsfbox{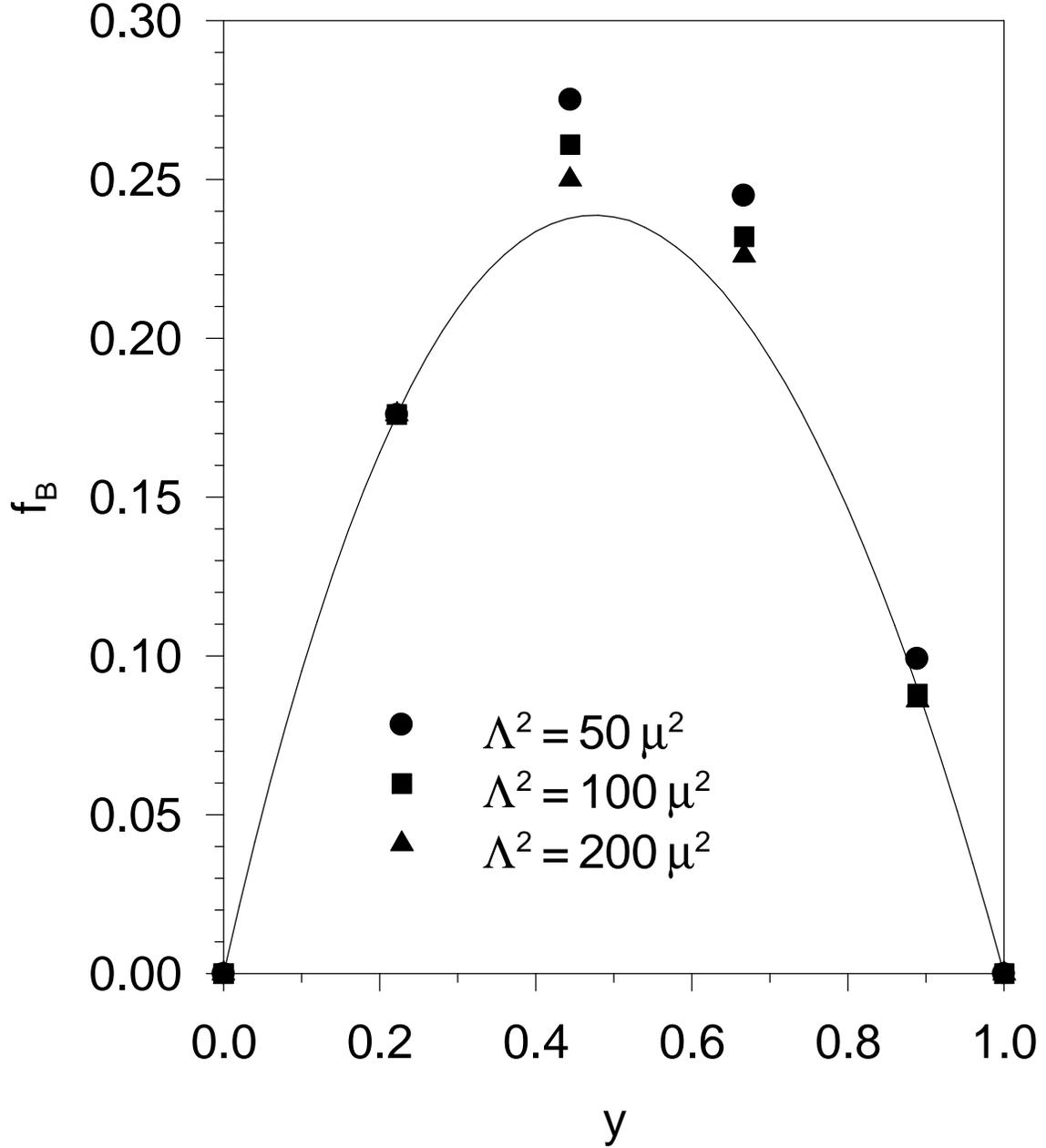}  }
\caption{\label{fig:fBLam-phi1}
The boson distribution function $f_B$ for different cutoff values,
with $\langle:\!\!\phi^2(0)\!\!:\rangle=1$ and numerical resolution
set at $K=9$ and $N_\perp=5$.  The solid line is the analytic result.}
\end{figure}

\begin{figure}
\centerline{\epsfxsize=\columnwidth \epsfbox{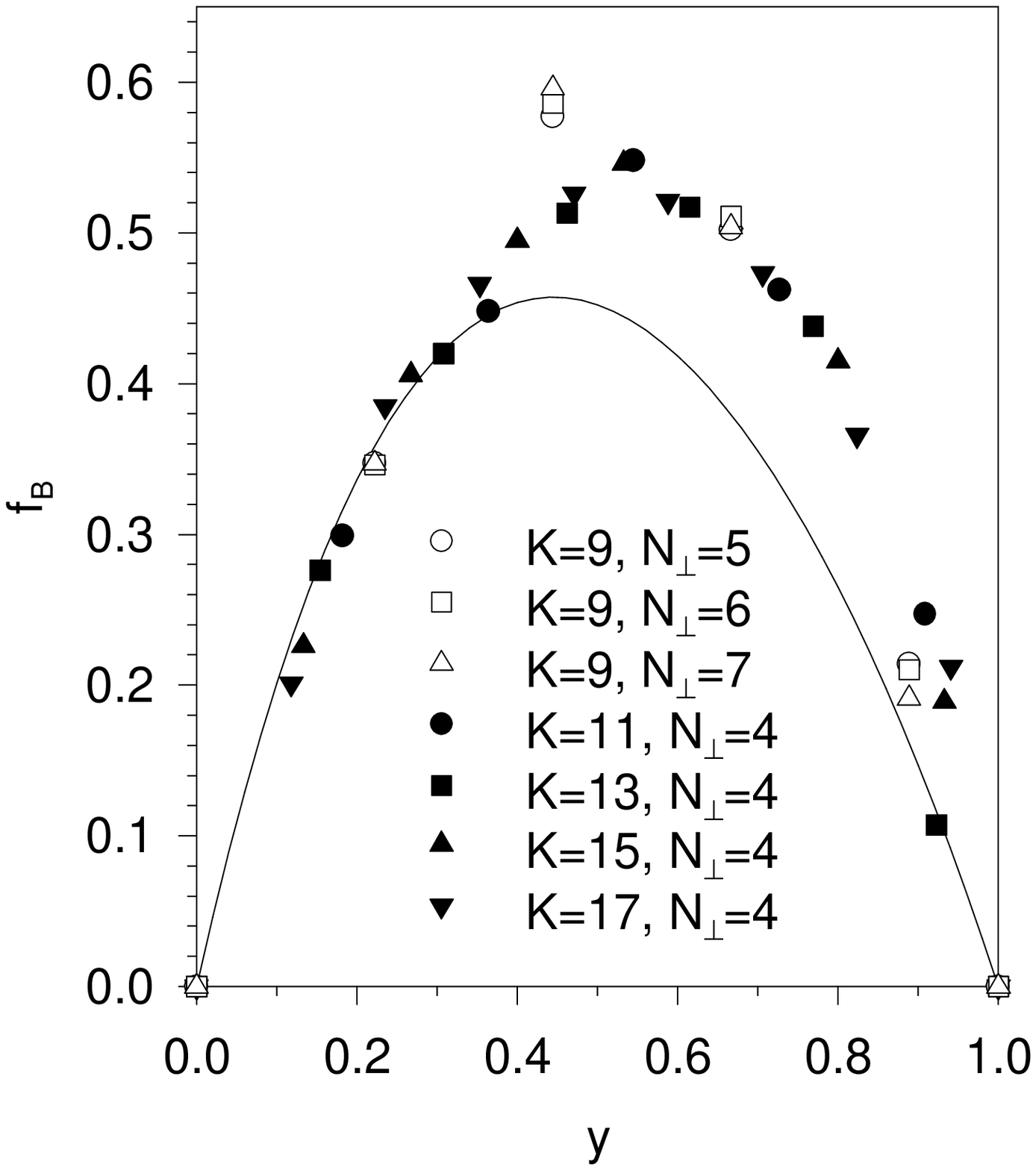}   }
\caption{\label{fig:fBNK-phi2}
Same as Fig.~\protect\ref{fig:fBNK-phi1} but for
$\langle:\!\!\phi^2(0)\!\!:\rangle=2$.}
\end{figure}

\begin{figure}
\centerline{\epsfxsize=\columnwidth \epsfbox{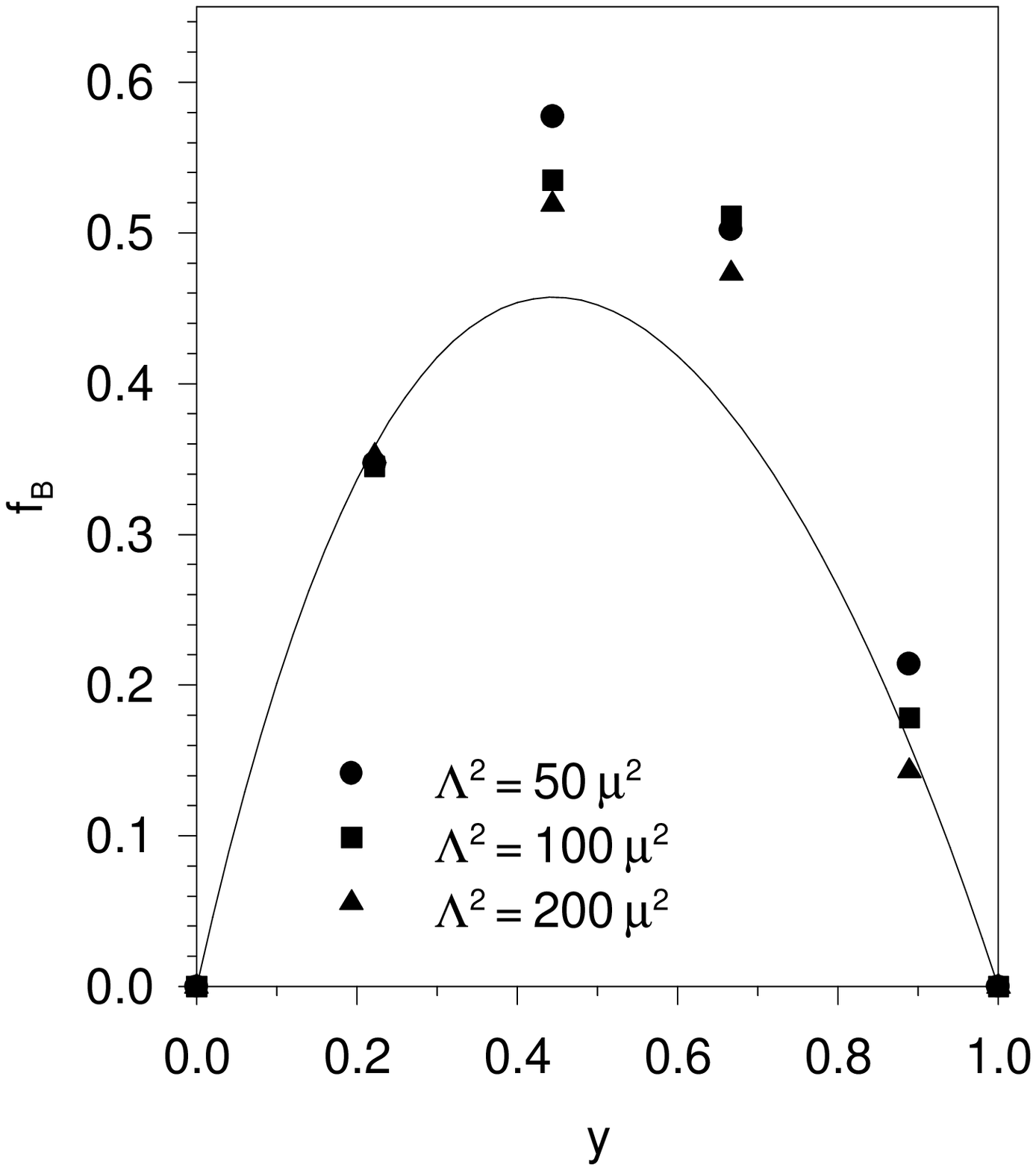}  }
\caption{\label{fig:fBLam-phi2}
Same as Fig.~\protect\ref{fig:fBLam-phi1} but for
$\langle:\!\!\phi^2(0)\!\!:\rangle=2$.}
\end{figure}

The results for $g$ and $\langle n_B\rangle$ are
surprisingly insensitive to variation in $K$ and $N_\perp$.
Only the cutoff $\Lambda^2$ is important.  This is mirrored
in the distribution functions shown in
Figs.~\ref{fig:fBNK-phi1}-\ref{fig:fBLam-phi2}.  Again,
variations in $K$ and $N_\perp$ make little difference;
however, one can see that the cutoff has an important
effect.  Smaller cutoffs produce an enhancement in the
interval $(0.4,\,0.8)$.\footnote{Recall that the distribution
function does not have a fixed normalization but instead
determines the average multiplicity, which is then also
enhanced at finite cutoff.}

\begin{figure}
\centerline{\epsfxsize=\columnwidth \epsfbox{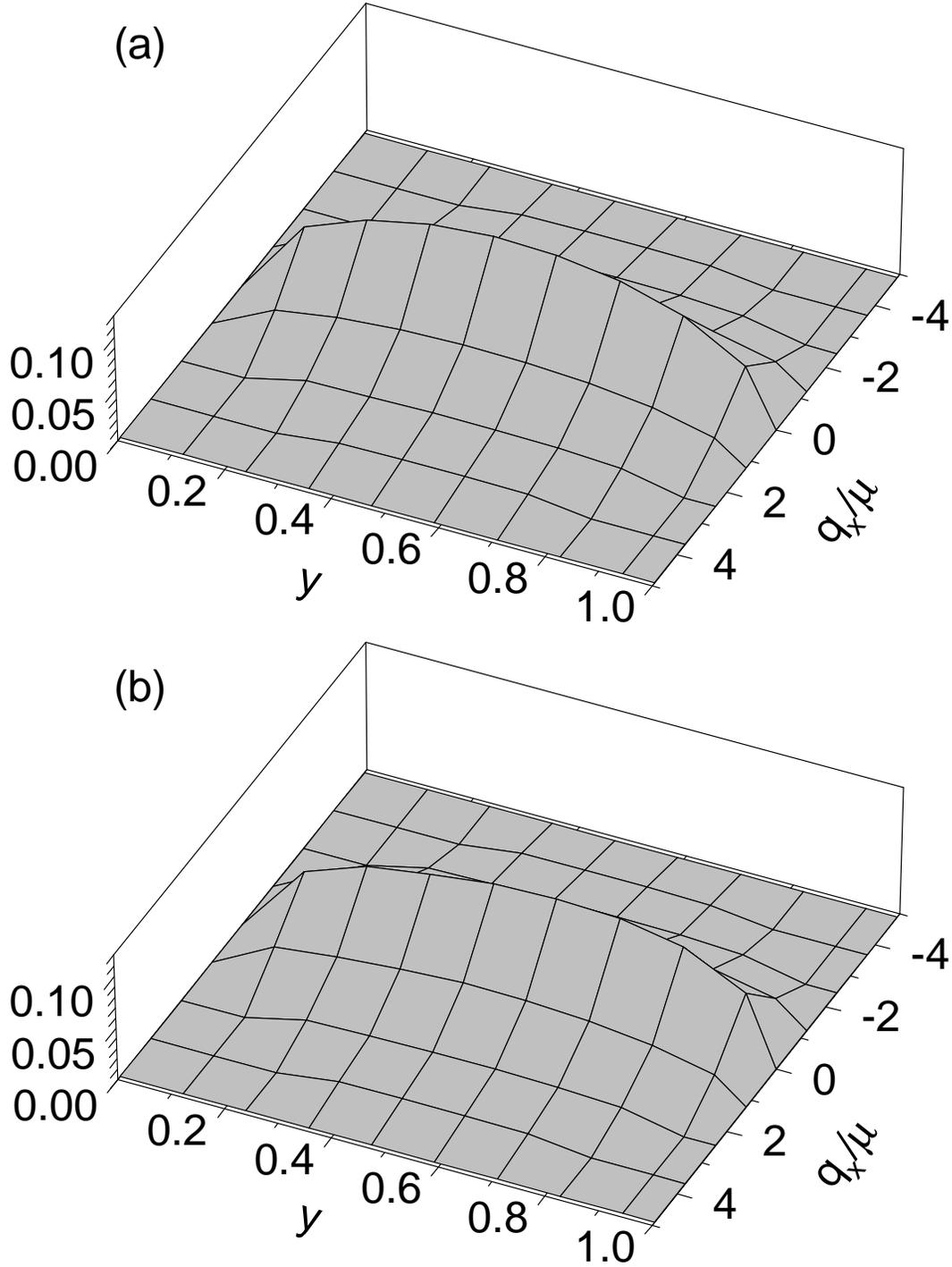}    }
\caption{\label{fig:OneBoson}
The one-boson amplitude $\psi^{(1,0)}$ as a function of longitudinal momentum
fraction $y$ and one transverse momentum component $q_x$ in the $q_y=0$ plane.
The analytic result is shown in (a) and the numerical result in (b) for
$\Lambda^2=50\mu^2$, $K=17$, and $N_\perp=4$.  Both correspond to
$\langle:\!\!\phi^2(0)\!\!:\rangle=1$.}
\end{figure}

The amplitude for the one-boson state is shown in Fig.~\ref{fig:OneBoson}.
The analytic result is shown for comparison.  As can be seen,
the two shapes are nearly identical.

\section{Conclusion}

In this paper we present a new method for the renormalization
of Hamiltonian light-cone-quantized field theories that maintains
Lorentz invariance and other symmetries.  The main difficulty
which is confronted by such methods is the construction of the
counterterms.  We employ the traditional
generalized Pauli--Villars method \cite{PauliVillars}.
With a sufficient number of Pauli--Villars fields, perturbation
theory is regulated while Lorentz symmetries and discrete
symmetries are preserved with a minimal number of counterterms.
These counterterms are generated automatically.  We hypothesize
that these counterterms are sufficient to regulate the nonperturbative
problem.

In Yukawa theory, Pauli--Villars regularization preserves
chiral symmetry \cite{ChangYan,BurkardtLangnau},
unlike the invariant-mass regulator.
A similar outcome arises in QED where one needs Pauli--Villars
regularization of the vacuum polarization loop to
recover its vanishing at $q^2\rightarrow 0$ \cite{BRS}.
These examples show that a covariant method is necessary in the
nonperturbative context to find all counterterms.

Given that the theory is finite by suitable Pauli--Villars
regularization, we can impose a regulator to limit the Fock
space so as to produce a tractable numerical problem.  A finite
matrix approximation can then be obtained with use of the DLCQ
procedure \cite{PauliBrodsky}.  In the finite matrix
problem we face the numerical difficulties of non-Hermitian
matrices and large basis sizes.  These difficulties are
successfully addressed in a $3+1$-dimensional
model \cite{SchweberGreenberg} constructed to have
an analytic solution.  This model requires one Pauli--Villars
boson as a regulator.  We also study the DLCQ approximation to
the one-loop fermion self energy in Yukawa theory, where three
Pauli--Villars bosons are needed \cite{ChangYan}.

The non-Hermitian matrices are handled by the complex
symmetric Lanczos diagonalization
algorithm \cite{Lanczos,Biorthogonal,Cullum}.  This technique
is ideal for the extraction of extreme eigenvalues and
their eigenvectors.  It takes full advantage of the sparsity
of the Hamiltonian matrix.  For a given basis size, storage
requirements are minimized.

The basis sizes required in the calculation are reasonable.
The presence of Pauli--Villars particles, at the chosen
mass and cutoff values, increased the model basis by only
100\% and the loop-calculation basis by 150\%.  Given
the sparsity of the matrix, increases of these magnitudes
are quite acceptable.  However, smooth convergence and
extrapolation from bases of minimal size require the
introduction of special integral weighting methods to
DLCQ.  The dramatic improvement which can occur is
illustrated in Fig.~\ref{fig:OneLoopSEcirc}.

With these methods we have obtained agreement between the
numerical and analytic solutions of our model.  The
convergence of the numerical result in longitudinal
and transverse resolution is remarkably rapid.
The result is sensitive only to the cutoff used to
limit the Fock space, but even there the convergence
to the analytic result is clear.  The methods seem
well suited to situations where low-mass states have
a small mean number of constituents.\cite{Hornbostel}

The natural next step is to extend the model
toward a more realistic theory, namely Yukawa theory.
The fermion can be given proper dynamics, and Yukawa-type
interactions can be reintroduced.  Once Yukawa theory
itself can be studied with our nonperturbative method,
there may be useful applications to the Higgs sector
of the Standard Model.  We are sufficiently encouraged by the
success of the Pauli--Villars program for the examples
discussed here to believe that
it will have general applicability to QCD in
$3+1$ dimensions.

\acknowledgments
This work was supported in part by the Minnesota Supercomputer Institute
through grants of computing time and by the Department of Energy.
The hospitality of the Telluride Summer
Research Center was also appreciated.

\appendix     

\section{Light-cone coordinates}
\label{sec:LCcoordinates}

We define light-cone coordinates\cite{Dirac} by
\begin{equation}
x^\pm \equiv x^0 \pm x^3 ,
\end{equation} 
with the transverse coordinates $\senk{x}\equiv (x^1,x^2)$ unchanged. 
Covariant four-vectors are written as \eg\ $x^\mu = (x^+,x^-,\senk{x}),$
with the spacetime metric
\begin{equation}
g^{\mu\nu}=\left(\matrix{0&2&0&0\cr
		   2&0&0&0\cr
		   0&0&-1&0\cr
		   0&0&0&-1\cr}\right) .
\end{equation}
Explicitly,
\begin{equation}
x\cdot y=g_{\mu\nu}x^\mu y^\nu = \ha(x^+y^-+x^-y^+)-\senk{x}\cdot\senk{y}.
\end{equation} 
We also make use of an underscore notation: for position-space variables we
write
\begin{equation}
\ub{x} \equiv (x^-,\senk{x}) ,
\end{equation} 
while for momentum-space variables
\begin{equation}
\ub{k} \equiv (k^+,\senk{k}) .
\end{equation} 
Then
\begin{equation}
\ub{k} \cdot\ub{x} \equiv\ha k^+x^- - \senk{k}\cdot\senk{x} .
\end{equation} 
Spatial derivatives are defined by
\begin{equation}
\del_+ \equiv {\del\over\del x^+}\ths,\qquad
\del_- \equiv {\del\over\del x^-}\ths,\qquad
\del_{i} \equiv {\del\over\del x^i} .
\end{equation} 

The gamma matrices $\g^\pm\equiv\g^0\pm\g^3=(\g^\mp)^\dagger$ satisfy the
familiar relation
\begin{equation}
\{\g^\mu,\g^\nu\}=2g^{\mu\nu}
\end{equation} 
with $g^{\mu\nu}$ the light-cone metric .  It is simple to verify
that the (Hermitian) matrices
\begin{equation}
\Lambda_\pm\equiv\ha \g^0\g^\pm
\end{equation} 
satisfy
\begin{equation}
\Lambda_\pm^2=\Lambda_\pm\
,\qquad\Lambda_\pm\Lambda_\mp=0\ths,
\qquad\Lambda_++\Lambda_-=1 ,
\end{equation} 
so that they serve as projectors on spinor space.  In the Dirac 
representation of the $\g$-matrices:
\begin{equation}
\Lambda_+=\ha\left(\matrix{1&0&1&0\cr 0&1&0&-1\cr
		      1&0&1&0\cr 0&-1&0&1\cr}\right) ,
\end{equation} 
which has two eigenvectors, both with eigenvalue $+1$:
\begin{equation}
\chi_{+\ha}={1\over\sqrt2}\left(\matrix{ 1\cr0\cr1\cr0\cr}\right)
\qquad\qquad
\chi_{-\ha}={1\over\sqrt2}\left(\matrix{ 0\cr1\cr0\cr-1\cr}\right) .
\end{equation} 
These serve as a convenient spinor basis for the expansion of the field
$\psi_+\equiv \Lambda_+\psi$ on the light cone.

\section{Weighting Methods}
\label{sec:WeightingMethods}

New weighting schemes have now been developed for use with the
DLCQ grid.  They are based on extensions of the trapezoidal
rule and Simpson's rule to the situation where the integration
domain does not end on a grid point; they are also related
to open Newton--Cotes formulas. The basic approach is to derive
formulas for one-dimensional integrals and then iterate them for
higher-dimensional integrals \cite{Luchini}.

\begin{figure}
\centerline{\epsfxsize=\columnwidth \epsfbox{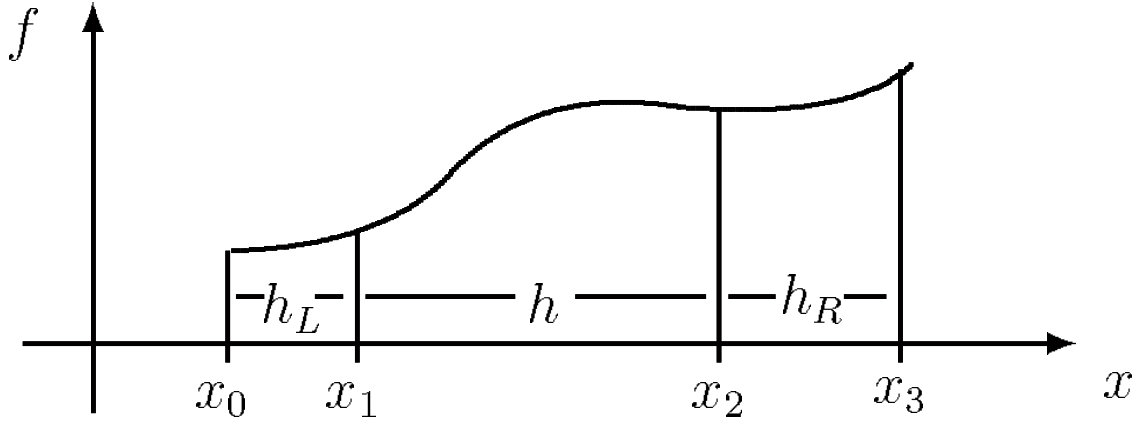} }
\caption{\label{fig:IntegrationRule}
Spacing of grid points for an arbitrary function.}
\end{figure}

The extended trapezoidal rule is obtained from consideration
of an integral from, say, $x_0$ to $x_3$.  The relevant graph is shown
in Fig.~\ref{fig:IntegrationRule}.  The grid points are at $x_1$
and $x_2$, which are separated by a standard spacing $h$.  The other
points are at the integration domain boundaries at distances of
$h_L$ and $h_R$ from the grid points.  The integral of a function
$f$ is then approximated by
\begin{equation} \label{eq:extendedTrap}
\int_{x_0}^{x_3}fdx\simeq a_1 f(x_1)+a_2 f(x_2)\,,
\end{equation}
with 
\begin{eqnarray} \label{eq:TrapCoefficients}
a_1&=&(h+h_L+h_R)(h+h_L-h_R)/2h\,,  \nonumber \\
a_2&=&(h+h_L+h_R)(h+h_R-h_L)/2h\,.
\end{eqnarray}
The coefficients $a_i$ are chosen to provide exact
results for linear functions.
The standard trapezoidal rule is recovered when $h_L=h_R=0$.
If $h_L=h_R=h$, a standard open Newton-Cotes formula results.
When the extended rule is combined with the standard 
rule for interior intervals,
a general composite rule is obtained.  The extended rule is
then used twice, once at each end, with $h_R$ or $h_L$ set to zero.

The extended Simpson's rule follows from similar steps.  Two
forms are needed, one for three grid points and another for four.
Any situation with more grid points can be handled with a
composite rule obtained by combining these rules with the
standard Simpson's rule.  For the three-point case, consider
an integral from $x_0$ to $x_4$, with grid points at $x_1$,
$x_2$, and $x_3$.  The regular grid spacing is $h$; the extra
points at the beginning and end are separated by $h_L$ and
$h_R$, respectively.  The approximation to the integral
is then
\begin{equation} \label{eq:extendedSimp3}
\int_{x_0}^{x_4} fdx\simeq
a_1 f(x_1)+a_2 f(x_2)+a_3 f(x_3)\,,
\end{equation}
where
\begin{eqnarray}
a_1&=&(4h^3+12h^2h_L+9hh_L^2+2h_L^3+3hh_R^2+2h_R^3)/12h^2\,, \nonumber \\
a_2&=&(4h^3-3hh_L^2-h_L^3-3hh_R^2-h_R^3)/3h^2\,, \\
a_3&=&(4h^3+12h^2h_R+9hh_R^2+2h_R^3+3hh_L^2+2h_L^3)/12h^2\,. \nonumber 
\end{eqnarray}
These coefficients are constructed to provide exact results for 
quadratic functions.  When $h_L=h_R=0$ they
reduce to the coefficients found in Simpson's rule, and, because
of the greater symmetry, the rule becomes exact for cubic functions
as well.  

The four-point rule is also exact for cubic functions.  It is
\begin{equation} \label{eq:extendedSimp4}
\int_{x_0}^{x_5} fdx\simeq
a_1 f(x_1)+a_2 f(x_2)+a_3 f(x_3)+a_4 f(x_4)\,,
\end{equation}
where
\widetext
\begin{eqnarray}
a_1&=&
 (9h^4+24h^3h_L+22h^2h_L^2+8hh_L^3+h_L^4-4h^2h_R^2-4hh_R^3-h_R^4)/24h^3\,,
          \nonumber \\
a_2&=&
 (27h^4-36h^2h_L^2-20hh_L^3-3h_L^4+18h^2h_R^2+16hh_R^3+3h_R^4)/24h^3\,, \\
a_3&=&
 (27h^4-36h^2h_R^2-20hh_R^3-3h_R^4+18h^2h_L^2+16hh_L^3+3h_L^4)/24h^3\,,
          \nonumber \\
a_4&=&
 (9h^4+24h^3h_R+22h^2h_R^2+8hh_R^3+h_R^4-4h^2h_L^2-4hh_L^3-h_L^4)/24h^3\,.
        \nonumber 
\end{eqnarray}
\narrowtext

\begin{figure}
\centerline{\epsfxsize=\columnwidth \epsfbox{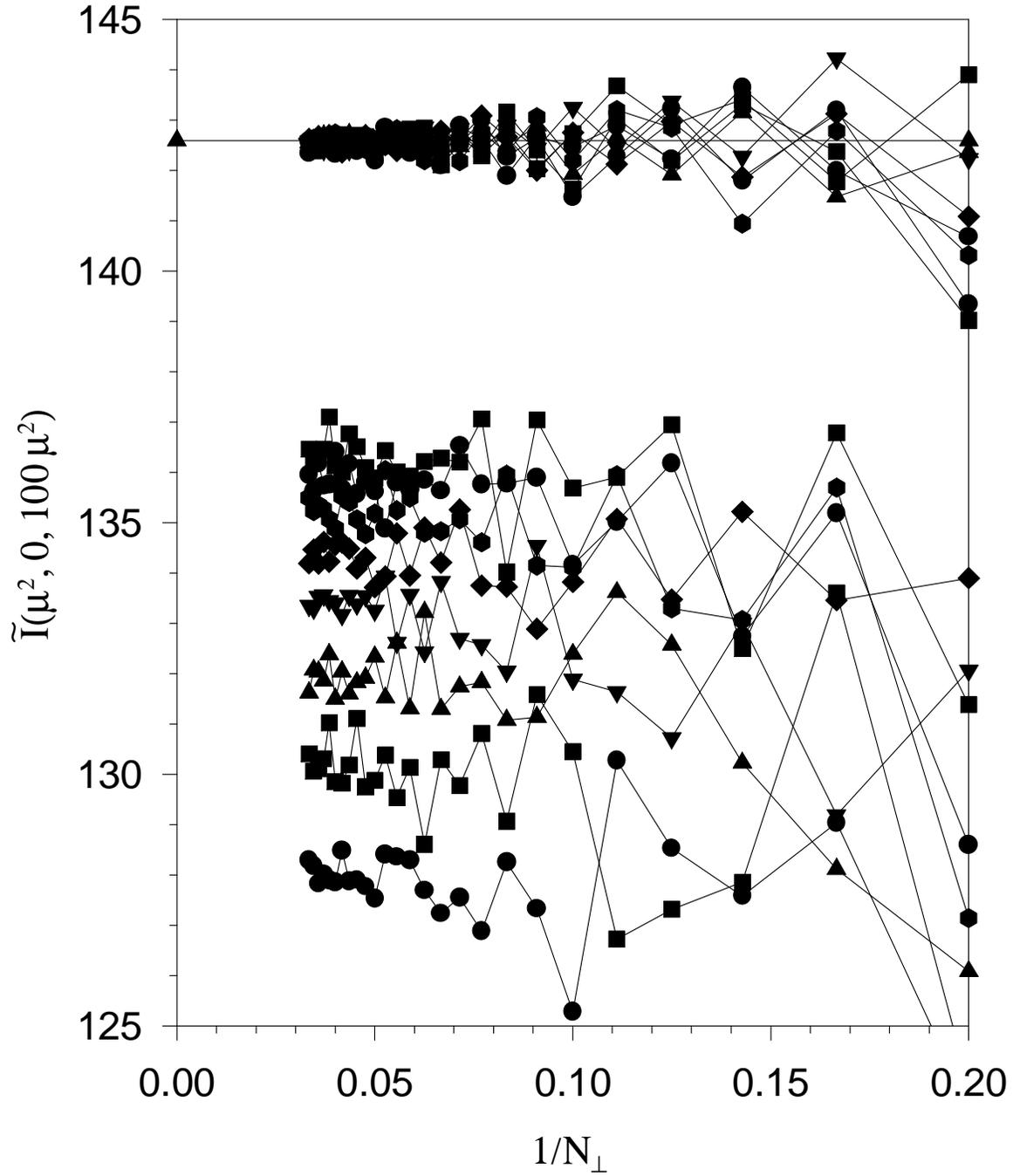} }
\caption{\label{fig:OneLoopSEtrap}
Same as Fig.~\protect\ref{fig:OneLoopSEcirc} except that the numerical
values oscillating about the analytic answer are computed
without transverse circular weighting and with only trapezoidal
weighting in the longitudinal and transverse directions.  The
lack of circular weighting destroys the smoothness of the results
shown in Fig.~\protect\ref{fig:OneLoopSEcirc}.}
\end{figure}

These integration formulas greatly reduce the size of the errors, 
as shown for the extended trapezoidal rule (\ref{eq:extendedTrap}) in
Fig.~\ref{fig:OneLoopSEtrap},
but they do not result in systematic behavior.  The lack of systematic
dependence is primarily due to the use of a square grid to
approximate a circular domain in the transverse direction.
One way of putting this is that the iterated
Cartesian integrals try to approximate
$\pi$ (a key factor in the area of the circle) as well as 
approximate the integral itself.

\begin{figure}
\centerline{\epsfxsize=\columnwidth \epsfbox{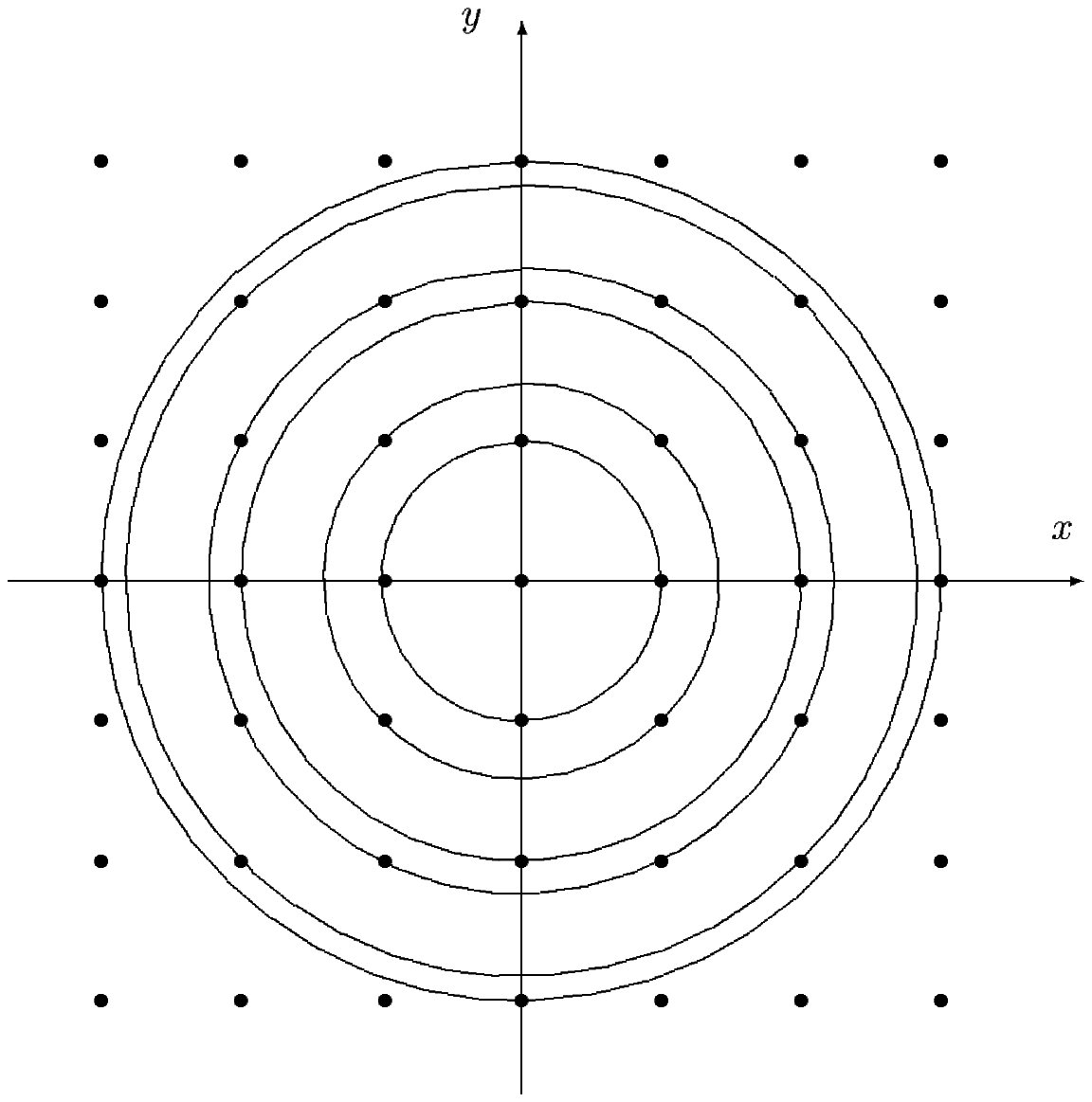} }
\caption{\label{fig:CircularGrid}
Square transverse grid with points on circles of varying radii.}
\end{figure}

To overcome this square/circle problem, the integral is
written in polar coordinates
\begin{equation}
\int dx dy f(x,y)=\frac{1}{2}\int_0^{2\pi}d\phi\int_0^{R^2}d(r^2)
                              \tilde{f}(r^2,\phi)\,.
\end{equation}
The points of the square grid lie on circles of varying radii $r_i$
shown in Fig.~\ref{fig:CircularGrid}.  The $r_i$ are easily computed
from the coordinates of the square grid.  The squares of these radii
are used as the grid points for a trapezoidal approximation to the
radial ($r^2$) integral.  Because the limit $R^2$ does not fall on
one of these points, the extended trapezoidal rule (\ref{eq:extendedTrap})
must be used for the last interval.  Clearly, the intervals are not
of equal length; however, they are on average of order $3h^2$, where
$h$ is the spacing in the square grid.  For the first 10 circles, the
average spacing in $r^2$ is actually closer to $2h^2$.

The number of points on the square grid that fall on any one circle come
in multiples of 4, because of reflection symmetries.  These points can be
used to approximate the angular integral on each circle via another
application of the trapezoidal rule.  For the self-energy integral $I$,
however, the integrand is independent of angle and one can simply use one
point or average the values at all points.  The contribution
to the weighting of a grid point is then the same for all
transverse points on the same circle.

\begin{figure}
\centerline{\epsfxsize=\columnwidth \epsfbox{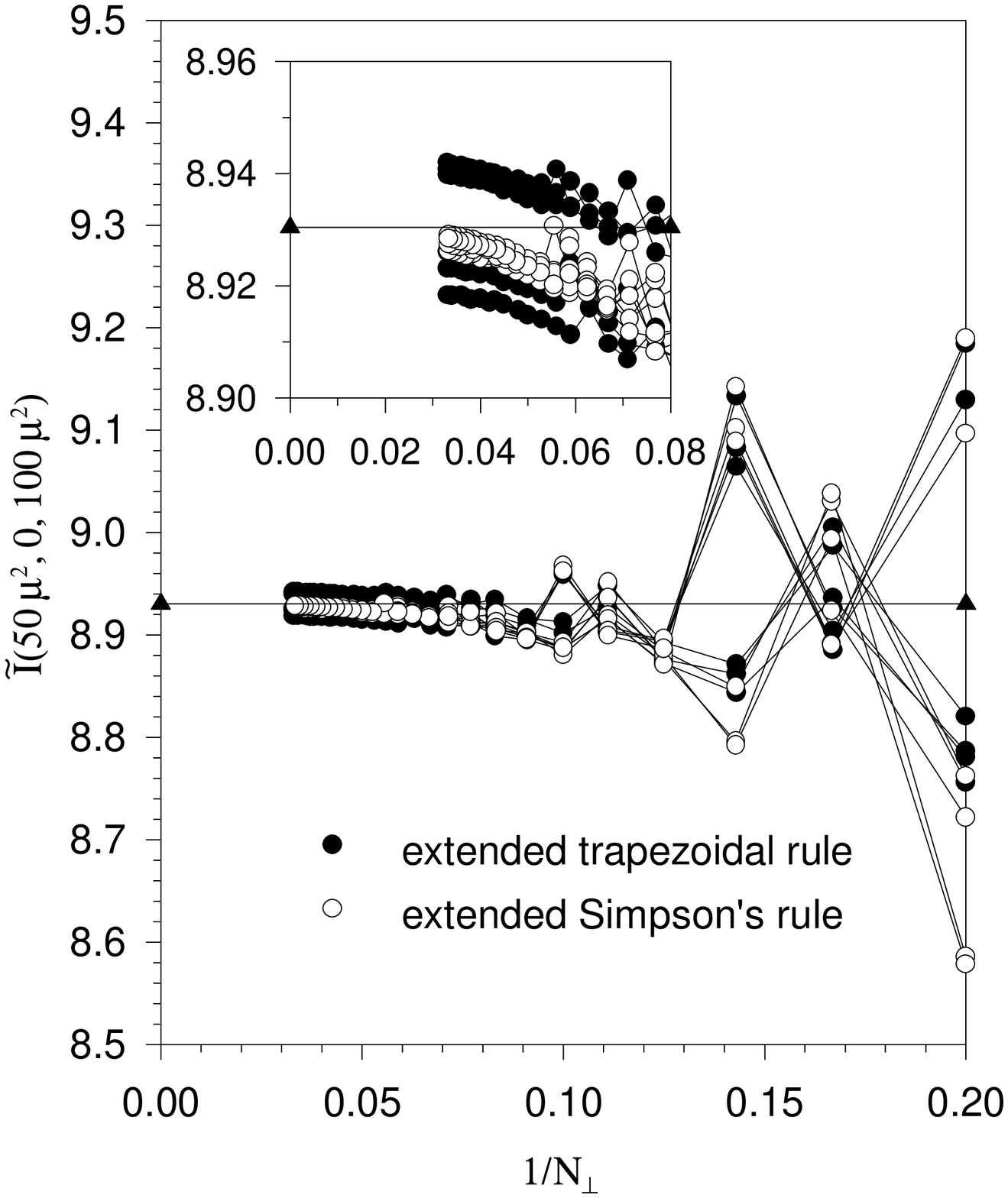} }
\caption{\label{fig:OneLoopSEsimp}
One-loop fermion self energy.  Results for
the extended trapezoidal and Simpson rules are compared. 
The horizontal line is the exact result.
The DLCQ grid parameters take the ranges $K=20,\,21,\ldots,25$ and
$N_\perp=5,\,6,\ldots,30$.
Points calculated with the same value of $K$ are connected by lines.}
\end{figure}

The circular weighting in the transverse direction can be combined
with either the extended trapezoidal rule or the extended
Simpson's rule in the longitudinal direction.  A comparison of the
two is shown in Fig.~\ref{fig:OneLoopSEsimp}.  The relatively large
excursions for small $N_\perp$ are due to the small number of grid
points involved for this case of a large boson mass.
For larger $N_\perp$ the extended Simpson's rule is seen to result 
in less excursion between different values
of $K$, and is preferred for the self-energy integral.
Results for the extended Simpson's rule are compared with
those of the ordinary DLCQ sum in Fig.~\ref{fig:OneLoopSEcirc};
in this case the results for the extended trapezoidal rule would
not be visibly worse.

\end{document}